\documentclass[apj,twocolumn]{openjournal}
\usepackage{natbib}
\usepackage{graphicx} 
\usepackage{xcolor}
\usepackage{tabularx}
\usepackage{textgreek}
\usepackage[utf8]{inputenc}
\usepackage[english]{babel}
\usepackage[caption=false]{subfig}
\UseRawInputEncoding
\usepackage{hyperref}
\hypersetup{
    unicode, 
    colorlinks=true,
    linkcolor=linkcolor,
    citecolor=linkcolor,
    filecolor=linkcolor,
    urlcolor=linkcolor,
}
\usepackage{color,colortbl}
\definecolor{linkcolor}{rgb}{0.0,0.3,0.5}
\usepackage{tensind}
\tensordelimiter{?}
\DeclareGraphicsExtensions{.bmp,.png,.jpg,.pdf}
\usepackage{verbatim}
\usepackage[normalem]{ulem}
\usepackage{orcidlink}
\usepackage{soul}

\usepackage[flushleft]{threeparttable}
\usepackage{booktabs}
\usepackage{amsmath}

\newcommand{\abcd}{{\tt abcd}}
\newcommand{\molh}{${\rm H_2}$}
\graphicspath{ {./figs/} }
\begin{document}

\title{Efficient semi-analytic modelling of Pop III star formation from Cosmic Dawn to Reionization \vspace{-15mm}}
\author{Sahil Hegde\orcidlink{0000-0002-9370-8061}$^{1\star\dagger}$}
\email{$\star$E-mail: sahil@astro.ucla.edu}
\thanks{$\dagger$ NSF Graduate Research \& NASA FINESST Fellow}
\author{Steven R. Furlanetto\orcidlink{0000-0002-0658-1243}$^1$}
\affiliation{$^1$Department of Physics \& Astronomy, University of California, Los Angeles, 475 Portola Plaza, Los Angeles, CA 90095, USA}

\shortauthors{\sc Hegde and Furlanetto}

\shorttitle{\sc Population III, as easy as \MakeLowercase{\abcd}}
% \shorttitle{\MakeLowercase{\tt abcd}:\sc A minimalist Pop III model}

\begin{abstract}
The quest to find the first stars has driven astronomers across cosmic time, from hopes to identify their signatures in their heyday at cosmic dawn to deep searches for their remnants in our local neighborhood. Such work crucially relies on robust theoretical modelling to understand when and where we expect pristine star formation to have occurred and survived. To that end, here we introduce a new {\tt a}nalytic {\tt b}athtub for {\tt c}osmic {\tt d}awn, the \abcd\ model, to efficiently trace the formation of the first stars from their birth through the first billion years of our universe's history, jointly following star formation out of pristine and metal-enriched gas over time. Informed by the latest theoretical developments in our understanding of star formation in molecular cooling halos, metal mixing, and early galaxies, we expand pre-existing minimal models for galaxy formation to include Population III stars and many of the processes --- both internal and environmental --- affecting their evolution, while remaining fast and interpretable. With this framework, we can bridge the gap between numerical simulations and previous semi-analytic models, as we self-consistently follow star formation in dark matter halos from the minihalo era through the epoch of reionization, finding that, under plausible physical conditions, pristine star formation can persist at a high level in the presence of Pop II star formation down to $z\sim 5$, but is limited to the most massive halos. We highlight areas of theoretical uncertainty in the physics underpinning Pop III star formation and demonstrate the effects of this uncertainty first on individual star formation histories and subsequently bracketing the range of global star formation levels we expect. Finally, we leverage this model to make preliminary observable predictions, generating forecasts for high-$z$ luminosity functions, transient rates, and the 21-cm global signal.
\end{abstract}


\maketitle

\section{Introduction}
In the canonical picture of structure formation in the universe, the first generations of stars formed in small dark matter (DM) `minihalos' tens of Myr after the Big Bang. Numerical simulations of this era --- wherein stars are born out of pristine hydrogen and helium clouds --- predict that the star formation process would have looked markedly different from what we see locally, resulting in a distinct early population of stars (for recent reviews, see \citealp{loeb_first_2013, bromm_formation_2013, klessen_first_2023}). While the theoretical understanding of their detailed properties and demographics remains uncertain, the consensus paradigm suggests that Pop III star formation would be a brief, bright burst in the earliest small gas clouds. Cooling in these clouds is driven by the comparatively inefficient transitions of molecular hydrogen (\molh), so Pop III stars are thought to form with a top-heavy initial mass function (IMF) and low star formation efficiency (SFE), with halos converting less than a few percent of their gas into a cluster dominated by stars with masses tens to hundreds of times that of our Sun \citep{abel_formation_2002, bromm_formation_2002, stacy_constraining_2013, hirano_one_2014, hirano_primordial_2015, jaura_trapping_2022, prole_fragmentation_2022, chon_impact_2024, sharda_population_2025, lake_stellar_2025}. In this picture, the first burst of star formation would rapidly enrich the surrounding medium and seed the next generation of metal-enriched, Pop II stars \citep{greif_simulations_2011, smith_first_2015, chiaki_metal-poor_2018, chiaki_seeding_2019, visbal_self-consistent_2020, feathers_global_2024, ventura_semi_2025}.

However, in recent years, there is an emerging understanding that pockets of pristine gas could survive hundreds of Myr beyond the era of the very first stars, as a product of inhomogeneous mixing of metals in the interstellar and circumgalactic media (ISM/CGM) of early galaxies (e.g., \citealp{pallottini_simulating_2014, sarmento_following_2017, sarmento_following_2018, sarmento_following_2019, liu_when_2020, venditti_first_2024, katz_challenges_2023, venditti_bursty_2025}), inefficient penetration of enriched bubbles into filaments in the intergalactic medium (IGM), or the delay of early star formation due to a background of \molh\ cooling-suppressive, `Lyman-Werner' radiation (e.g., \citealp{johnson_first_2013, xu_late_2016, xu_xray_2016, wolcott-green_h2_2019, skinner_cradles_2020, hegde_self-consistent_2023}). In such contexts, pristine star formation could persist not only in the smallest halos, but also in more massive, `atomic-cooling' halos ($T_{\rm vir} \gtrsim 10^4 K$) well into the epoch of reionization, in principle providing a more observationally accessible probe of pristine star formation (e.g., \citealp{xu_late_2016, riaz_unveiling_2022, venditti_needle_2023, venditti_first_2024}).

On the observational front, the search for the first stars has been a longstanding quest in modern astronomy. With increasing instrument sensitivity and creative observing strategies, for nearly three decades astronomers have doggedly sought out these seeds of the earliest luminous structures both near and far. While such searches have yielded promising candidates over the years, none have yet been convincingly confirmed as detections of the very first stars (for some recent such efforts, see e.g., \citealp{vanzella_candidate_2020, vanzella_extremely_2023, wang_strong_2024, maiolino_possible_2024, chiti_detailed_2023, chiti_enrichment_2024, fujimoto_glimpse_2025, morishita_pristine_2025, cai_metal_2025}). Along with the absence (or weakness) of metal lines, these searches typically rely on one or more of the prototypical `smoking gun' signatures of star formation with a top-heavy IMF, such as strong, high-ionization emission lines produced by hydrogen and helium, detections of dominant nebular continua, or atypical chemical abundance patterns that may reflect relics of the Pop III era. However, at some level, these discoveries rely on serendipity, and if successful, will still be challenging to interpret in the context of population demographics due to necessarily small samples. Even if pristine star formation were to persist to late times, where direct observations become more feasible, it is unclear just how abundant these star formation sites would be. As a result, there are a number of probes attempting to detect the imprints of early star formation in the intergalactic gas --- through metrics such as 21-cm or emission line intensity mapping --- a regime in which a \textit{statistical} understanding of early star forming populations becomes more achievable (see e.g., \citealp{mirocha_unique_2018, mebane_effects_2020, sun_revealing_2021, magg_effect_2022, munoz_impact_2022, parsons_probing_2022, ventura_role_2023, hegde_self-consistent_2023, cruz_effective_2025, ventura_semi_2025, gessey-jones_impact_2022, gessey-jones_determination_2025} for some recent investigations). Given the relative dearth of observational constraints at the earliest cosmic times --- and even into the epoch of reionization --- all such observations are heavily guided by theoretical models, both numerical and analytic, of early star and galaxy formation. In the former aforementioned regime (in direct searches for Pop III systems), theoretical estimates are crucial to optimize survey configurations and temper expectations, while in the latter (intensity mapping surveys), they will also be necessary to interpret measurements and infer physical parameters.

To this end, comprehensive theoretical models are vital and, given the broad parameter space, efficiency of such models is key. Thus, in this work, we present a new, semi-analytic description of early star formation that tracks halos from the birth of their first stars out to the end of reionization, incorporating many of the relevant physical mechanisms influencing these galaxies. By emphasizing simplicity in this model, we make evident the effects of the individual physical components and provide a framework that can be \textit{efficiently} leveraged to guide observations and carry out parameter inference from data.

To summarize our approach, we begin with the baseline `minimalist' galaxy model first outlined in \cite{furlanetto_minimalist_2017} and extended to include bursty star formation in \cite{furlanetto_bursty_2022}. These models use a series of simple evolution equations to characterize the dominant forces governing the first phases of galaxy formation and evolution (for other work employing a framework that is similar in spirit, see e.g., \citealp{bouche_impact_2010, dave_analytic_2012, dekel_toy_2013, lilly_gas_2013, mirocha_effects_2020, kravtsov_grumpy_2022}). \cite{furlanetto_minimalist_2017} demonstrated that tracking the growth of a gas reservoir $m_g$ that is converted into an evolving stellar population $m_\star$, with the interplay between the two regulated by mechanical stellar feedback prescriptions, provides a reasonable fit to galaxy luminosity functions from $z\sim 6-10$ (both pre- and post-JWST). 

In this work, in order to better describe galaxy populations at earlier times, while maintaining the spirit of interpretability, flexibility, and efficiency, we couple that framework with the Pop III model introduced in \cite{mebane_persistence_2018} and \cite{hegde_self-consistent_2023} to account for the evolution of galaxies from the very first star-forming clouds born out of pristine gas to their transition to hosting metal-enriched star-forming disks at later times, building a an analytic bathtub for cosmic dawn, the \abcd\ model.\footnote{We implement this model in a Python package of the same name, hosted at https://github.com/hegdesahil/abcd.} The key new feature of the \abcd\ model is that we evolve these two populations in a two-component `bathtub' of sorts, treating the pristine and enriched gas reservoirs as separate buckets from which stars are born. Because the relative rates of star formation in these two phases will be highly sensitive to the segregation of gas between the two reservoirs, we incorporate a more sophisticated treatment of gas cycling in the halo than required by previous iterations of the minimalist model. This is manifested in the inclusion of a circumgalactic medium (CGM; inspired by the work of \citealp{carr_regulation_2023, pandya_unified_2023, voit_equilibrium_2024a, voit_equilibrium_2024b}), to which we defer the process of metal pollution. We then self-consistently couple that expanded galaxy model with an approximate treatment of the global state of the intergalactic medium in order to estimate the effects of reionization and metal enrichment on star formation in individual systems.

In what follows, we first introduce the \cite{furlanetto_bursty_2022} evolution equations to contextualize the baseline model (Section~\ref{sec:burst_baseline}) then describe the new physical processes we introduce, first those internal to the galaxy (Section~\ref{sec:internal_modifications}) and then those set by the evolution of the IGM (Section~\ref{sec:external_modifications}). For a summary of the most salient effects in these sections and a schematic overview of the model, we direct the reader to Sections~\ref{sec:internal_summary}, and \ref{sec:external_summary}, and Figure~\ref{fig:model_schematic}, respectively. We then integrate these histories over a population of galaxies and examine the resulting trends in Section~\ref{sec:pop_quantities}. We situate our model in the context of previous work in Section~\ref{sec:past_work}, sketch out some observational estimates that this model naturally produces (Section~\ref{sec:observables}), and, finally, in Section~\ref{sec:conclusion}, we conclude.

Throughout this work, we use a flat $\Lambda$CDM cosmology with $\Omega_{\rm m} = 0.3111$, $\Omega_{\Lambda} = 0.6889$, $\Omega_{\rm b} = 0.0489$, $\sigma_8 = 0.8102$, $n_s = 0.9665$, and $h = 0.6766$, consistent with the results of \citet{Planck21}

\section{The baseline burst model}\label{sec:burst_baseline}
In this section, we summarize the key components of the original \cite{furlanetto_bursty_2022} model for bursty star formation in high-$z$ galaxies driven by delayed stellar feedback.

\subsection{Halo assembly}
As in the original minimalist model, we model halo growth by abundance matching the halo mass function across redshift --- i.e., we assume that halos maintain a constant number density over time. This assumption results in halos that grow smoothly over time with rates similar to the analytical expressions used to model the results of numerical simulations, such as that presented in \cite{dekel_toy_2013}, $\dot{m}_h = Am_h(1+z)^{5/2}$. Specifically, we derive the amount that a halo of mass $m_1$ will have grown between redshifts $z_1$ and $z_2$ by solving for $m_2$ in the following:
\begin{equation}
    \int_{m_1}^\infty n(m,z_1) dm = \int_{m_2}^\infty n(m,z_2) dm. 
\end{equation}
\cite{mirocha_importance_2021} demonstrated that the assembly histories that result from this assumption provide a good match to the average growth rates of high-redshift halos in numerical simulations, which include, e.g., the contribution of mergers. Following that work, we fiducially choose the \cite{trac_scorch_2015} halo mass function to characterize our halo growth histories.

With the halo growth rate inferred in this manner, the total gas accretion rate onto the galaxy is $\dot{m}_{\rm g, acc} = f_gf_b\dot{m}_h$, where $f_b \equiv \Omega_b/\Omega_m$ and the fraction of gas eligible to fall into a halo is described in detail in Section~\ref{ssec:reion_feedback}.

\subsection{The evolution equations}
Having derived the gas accretion rates this way, a Pop II galaxy evolves subject to the following evolution equations. 
\begin{align}
    \dot{m}_g &= \dot{m}_{\rm g, acc} - \dot{m}_\star - \dot{m}_w \\
\dot{m}_\star &= \frac{\epsilon_{\rm ff}}{t_{\rm ff}} m_g\bigg|_{\Sigma_g > \Sigma_{\rm crit}} \\
\dot{m}_w &= \eta \dot{m}_\star^{\mathcal{D}} \\
\dot{m}_Z &= (y_Z - Z)\dot{m}_\star-\eta Z \dot{m}_\star^{\mathcal{D}}
\end{align}
Note here and in subsequent equations that a superscript $\mathcal{D}$ is applied to a term only when we are evaluating that particular quantity at an earlier time, as is discussed in more detail below. 

These equations can be interpreted fairly straightforwardly: gas accretes onto the halo following the cosmological halo mass accretion rate and settles into a disk. This gas will then be eligible for star formation once the disk exceeds a critical surface density (set by the Toomre criterion $Q\sim 1$, which characterizes gravitational instability)
\begin{equation}\label{eq:crit_disk_dens}
    \Sigma_{g, \rm crit}\sim \frac{c_{\rm eff}\Omega}{G},
\end{equation}
where $\Omega = v_{\rm c}/2\pi r_{\rm d}$ is the orbital frequency and the effective sound speed, which is related to the gas fraction and scale height of the disk, is assumed to be $c_{\rm eff}\sim 0.1v_{\rm c}$ \citep{liu_effects_2024}. To estimate the galaxy's free-fall time, we follow the calculation outlined in Section 3.3 of \cite{furlanetto_bursty_2022} (and the references therein), namely taking $t_{\rm ff}\sim 0.2t_{\rm orb}$. This criterion assumes that the disk follows an isothermal potential and is motivated by the disk models described in \cite{faucher-giguere_feedback_2013} and \cite{faucher-giguere_model_2018} through \cite{liu_effects_2024}. Star formation then proceeds with an efficiency per free-fall time $\epsilon_{\rm ff}$ until stellar feedback-driven winds evacuate the gas reservoir after a delay time $\mathcal{D}$, set by the lifetimes of massive stars ($t_{\rm short}=5\ {\rm Myr}, t_{\rm long} = 30\ {\rm Myr}$).\footnote{In practice, the delayed SFR is given by the average SFR over a window of massive stellar lifetimes $t_{\rm life}\sim (t_{\rm short}, t_{\rm long})$: $\dot{m}_\star^\mathcal{D}(t) = \frac{1}{t_{\rm long} - t_{\rm short}}\int_{t-t_{\rm long}}^{t-t_{\rm short}} \dot{m}_\star(t') dt'$} While we calibrate the choice of $\epsilon_{\rm ff}$ in the baseline, Pop II model to mimic the results of numerical simulations, we later estimate this quantity by accounting for other, instantaneous stellar feedback sources, such as radiation pressure, when expanding this framework into the zero-metallicity regime (Section~\ref{sec:rad_feedback}). The strength of these winds is characterized by the feedback mass-loading parameter $\eta(m_h, z)$, for which the exact scaling varies based on the assumed feedback mechanism, but takes a generic form \citep{furlanetto_minimalist_2017}
\begin{equation}\label{eq:minimalist_eta}
    \eta(m_h, z) = C\bigg(\frac{10^{11.5} M_\odot}{m_h}\bigg)^\xi \bigg(\frac{9}{1+z}\bigg)^\sigma.
\end{equation}
For example, for energy-(momentum-) regulated feedback, the mass loading parameter varies with halo mass as $\xi = 2/3$ (1/3) and with redshift as $\sigma = 1$ (1/2), and the normalization $C$ is largely set by the IMF. This framework results in a series of star formation cycles, or `bursts,' which gradually damp out as the halo (and thus its associated escape velocity) grows.  

In tandem, these star formation episodes pollute the ISM with metals (following an assumed yield $y_Z$), resulting in an associated evolution of the galaxy's metallicity $Z\equiv m_Z/m_g$.

This framework has been successfully applied to model galaxy populations at $z\sim 6-10$ in a variety of contexts (see e.g., \citealp{furlanetto_bursty_2022, pallottini_mass_2025, liu_effects_2024}), but, as written, is only equipped to model the evolution of a single, homogeneous stellar population, a requirement that we relax in the following sections.

\section{Internal evolution}\label{sec:internal_modifications}
Building from the minimal burst model summarized in Section~\ref{sec:burst_baseline}, in this section we describe the components of the \abcd\ model \textit{internal to the host DM halo} and their effects on the resulting galaxy's evolution.

\begin{figure*}
    \centering
    \includegraphics[width=0.9\linewidth]{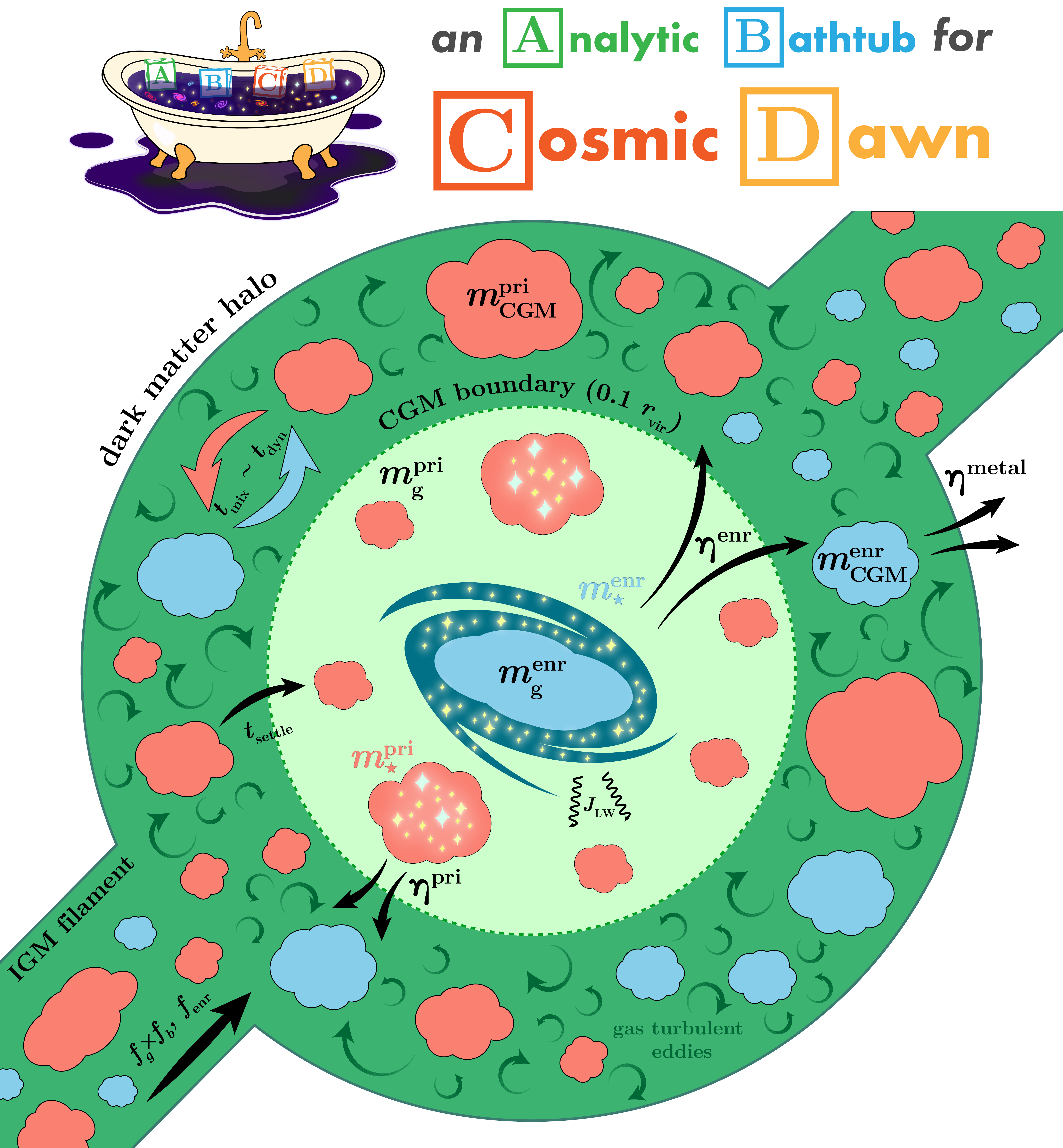}
    \caption{A schematic overview of the various components of the \abcd\ model. As in the subsequent plots, DM reservoirs are represented in green, pristine reservoirs in red, and enriched reservoirs in blue. While in detail the model does \textit{not} contain any spatial information about the star-forming and gaseous components of the different mass reservoirs, this schematic serves to illustrate the rough picture motivating the choices of the different quantities and parameterizations underpinning the model. Most of the quantities referenced in this schematic are specified in Table~\ref{tab:constants} or in eqs.~\ref{eq:m_ism_ev}-\ref{eq:t_mix}. As noted in those equations and the table, the $m$ quantities correspond to masses in different reservoirs, $t$ to the timescales over which the different processes proceed, and $\eta$ to the mass loading factors computed for feedback and metal ejection (eqs.~\ref{eq:minimalist_eta} and \ref{eq:eta_metal}). A broad overview of the model is as follows: pristine material flows into the circumgalactic region of a DM halo along filaments from the IGM (red clouds). After some time ($\mathcal{O}(t_{\rm dyn})$), this gas settles into the central ISM and eventually forms stars, driving the buildup of pristine stellar mass $m_\star^{\rm pri}$. Once these stars die, they eject metal-enriched gas into the circum- and intergalactic media around that halo (with strengths characterized by the mass-loading parameters $\eta^{\rm pri}$, $\eta^{\rm metal}$), kick-starting the local and global metal enrichment process. As the enriched reservoir (blue clouds) builds up, it too settles into the ISM, from which a central disk forms and Population II stars are born ($m_\star^{\rm enr}$). These cycles proceed in both reservoirs --- enriched and pristine --- as the halo's local (and the IGM) metallicity slowly grows (which modifies $f_{\rm enr}$; Section~\ref{sec:global_enrichment}), subject to the additional effects of reionization feedback (which alters $f_{g}$; Section~\ref{ssec:reion_feedback}).}
    \label{fig:model_schematic}
\end{figure*}

\subsection{The dual ISM}\label{sec:two_phase_ISM}
In order to extend this model to the earliest stages of a galaxy's evolution, we must account for the existence of and transition between two regimes of star formation, distinguished by the presence of metals or lack thereof. To do this, we introduce an analogous set of reservoirs to those outlined in Section~\ref{sec:burst_baseline} corresponding to star formation in a pristine, `Pop III' phase.\footnote{In this text, we will refer to pristine and Pop III star formation interchangeably, though we note that in some definitions, the term Population III is used exclusively to refer to the \textit{first} generation of stars hosted in a halo. In such instances, it can also be common to refer to Pop III star formation as occurring below and sensitive to a chosen critical metallicity, which we circumvent with the dual ISM framework.} In other words, we bifurcate the ISM into two independent, yet mostly-identical gas and stellar reservoirs satisfying the following set of evolution equations
\begin{align}\label{eq:m_ism_ev}
\dot{m}_g^{\rm pri} &= \dot{m}_{\rm settle}^{\rm pri} - \dot{m}_\star^{\rm pri} - \dot{m}_w^{\rm pri} \\
\dot{m}_\star^{\rm pri} &= \frac{\epsilon_{\rm ff}^{\rm pri}}{t_{\rm ff}} m_g^{\rm pri} \bigg|_{m_h > m_{\rm crit, H_2},\ m_g^{\rm pri} > m_{J}}\\
\dot{m}_w^{\rm pri} &= \eta^{\rm pri} \dot{m}_\star^{\rm pri, \mathcal{D}}\\
\dot{m}_g^{\rm enr} &= \dot{m}_{\rm settle}^{\rm enr} - \dot{m}_\star^{\rm enr} - \dot{m}_w^{\rm enr}\\
\dot{m}_\star^{\rm enr} &= \frac{\epsilon_{\rm ff}^{\rm enr}}{t_{\rm ff}} m_g^{\rm enr}\bigg|_{\Sigma_g^{\rm enr} > \Sigma_{\rm g, crit}} \\
\dot{m}_w^{\rm enr} &= \eta^{\rm enr} \dot{m}_\star^{\rm enr, \mathcal{D}} \\
\dot{m}_Z^{\rm gal} &= \dot{m}_{\rm settle}^{Z} + (y_Z^{\rm enr} - Z)\dot{m}_\star^{\rm enr}-\eta^{\rm enr} Z \dot{m}_\star^{\rm enr,\mathcal{D}},
\end{align}
where we have now added the superscripts `pri' and `enr' to denote the pristine and enriched components, respectively. Such a picture --- where we treat the two sets of reservoirs as independent --- is motivated by recent simulations, which argue that inhomogeneous mixing of metals can allow for continued inflow of pristine material and star formation within those clumps out to late stages in a galaxy's evolution (see e.g., \citealp{pallottini_simulating_2014, pan_modeling_2013, sarmento_following_2017, sarmento_following_2018, sarmento_following_2019, venditti_needle_2023, venditti_first_2024, venditti_hide_2024}). In essence, here we are imagining two spatially segregated components of star-forming gas within the galaxy: a larger, star-forming disk consistent with the conventional picture of Pop II star formation in a relatively mature galaxy (i.e., the model of Section~\ref{sec:burst_baseline}, where metals have been efficiently mixed), and a collection of pristine clumps fed by filaments (into which metals carried by stellar winds and supernovae do not efficiently penetrate) --- see Figure~\ref{fig:model_schematic} for a visualization of this structure. Note, however, that we do not track the spatial locations of these two components and instead rely on average quantities informed by this picture to model the system. In practice, we separate these phases by introducing two independent and as-yet undefined source terms, $\dot{m}_{\rm settle}^{\rm (i)}$. 

For reference, the constants chosen in these evolution equations and the following are summarized in Table~\ref{tab:constants}.

\begingroup 
    \setlength{\tabcolsep}{10pt} 
    \renewcommand{\arraystretch}{1.5} 
    \setlength\extrarowheight{2pt}
    \begin{table*}
        \caption{{Parameters chosen in the fiducial model.}}\label{tab:constants}
        \centering
        \begin{tabular}{c | c c }  
            \hline \hline 
            Parameter(s) & Description & Fiducial value \\
            \hline 
            $R_{\rm ISM}$ & radius of the ISM-CGM boundary & $0.1 R_{\rm vir}$ \citep{pandya_unified_2023} \\
            $\zeta, \alpha_{\rm vbc}, f_X, v_{\rm bc}$ & $m_{\rm crit}$ parameters (see \citealp{hegde_self-consistent_2023}) & $0.16, 5, 10, 1\sigma_{\rm vbc}$ \\
            $\epsilon_{\rm ff}^{\rm pri}$ & Pop III star formation efficiency per free fall time  & see Sections~\ref{sec:rad_feedback} and \ref{sec:early_SFRD} (calibrated to 0.001)\\
            $\epsilon_{\rm ff}^{\rm enr}$ & Pop II star formation efficiency per free fall time  & 0.015 \\
            $C^{\rm pri},\xi^{\rm pri},\sigma^{\rm pri}$ & Pop III SN feedback parameters (eq.~\ref{eq:minimalist_eta}) & 1, 0.66, 1 \\
            $C^{\rm enr},\xi^{\rm enr},\sigma^{\rm enr}$ & Pop II SN feedback parameters (eq.~\ref{eq:minimalist_eta}) & 0.08, 0.75, 0 \\
            $\alpha, m_{\rm char}, \beta$ & Pop III IMF parameters (eq.~\ref{eq:IMF}) & 2.35, $20\ M_\odot$, 1.6 \\
            $f_w$ & fraction of SN energy available to drive winds & 0.3 (calibrated to $Q_{\rm IGM}$, Section~\ref{sec:global_enrichment}) \\
            $\mathcal{D}_{\rm pri}$ & delay time for Pop III SN feedback & $\langle t_{\rm life}\rangle_{\rm IMF}\sim 7\ {\rm Myr}$ \\
            $\mathcal{D}_{\rm enr}$ & delay time for Pop II SN feedback & $5-30\ {\rm Myr}$ \\
            $f_{\rm esc}^{\rm pri}$ & escape fraction of ionizing photons for Pop III stars & 0.1 \\
            $f_{\rm esc}^{\rm enr}$ & escape fraction of ionizing photons for Pop II stars & 0.1 \\
            \hline \hline  
        \end{tabular}  
        \label{table:table}
    \end{table*}
\endgroup

\subsection{The circumgalactic medium}\label{sec:cgm}
Evidently, the process of mixing is a key feature of the model; in order to robustly understand the relative rates of star formation in the two phases, we need to understand the interplay and transfer of material between the two reservoirs. To this end, we extend the baseline model framework to include a circumgalactic medium (CGM) component (with two analogous phases) in a manner inspired by the minimal CGM model introduced by \cite{pandya_unified_2023}. 

In particular, we introduce an additional set of evolution equations describing the halo of gas surrounding the central galaxy into which material flows, both cosmologically, from the IGM, and driven by stellar feedback processes out of the ISM. This, therefore, serves as the natural interface through which we couple our reservoirs, as follows:
\begin{align}\label{eq:CGM_pri}
\dot{m}_{\rm CGM}^{\rm pri} ={}& (1-f_{\rm enr})\dot{m}_{\rm g, acc} -\dot{m}_{\rm settle}^{\rm pri} - \dot{m}_{\rm esc}^{\rm pri} - \frac{m_{\rm pri}}{t_{\rm mix}}\\
\begin{split}\label{eq:CGM_enr}
     \dot{m}_{\rm CGM}^{\rm enr} ={}& f_{\rm enr}\dot{m}_{\rm g, acc} +\dot{m}_{w}^{\rm pri, \mathcal{D}} + \dot{m}_{w}^{\rm enr, \mathcal{D}}-\dot{m}_{\rm settle}^{\rm enr} \\ 
     &- \dot{m}_{\rm esc}^{\rm enr} + \frac{m_{\rm pri}}{t_{\rm mix}}
\end{split}\\
\begin{split}
   \dot{m}_{Z, \rm CGM} ={}& Z_{\rm IGM}f_{\rm enr}\dot{m}_{\rm g, acc} + y_Z^{\rm pri}\dot{m}_{w}^{\rm pri} + Z\dot{m}_{w}^{\rm enr}\\  &- \dot{m}_{\rm settle}^{Z} - Z_{\rm CGM}\dot{m}_{\rm esc}^{\rm enr}\label{eq:CGM_Z}
\end{split}
\end{align}
Here we have introduced a coefficient $f_{\rm enr}$ to describe the fraction of material entering the halo that is efficiently mixed with metals (for now taken to be zero) and segregates material between the two parent CGM reservoirs. While the pristine reservoir here is only fueled cosmologically, the enriched reservoir also grows as enriched outflows evacuate material from the ISM. In other words, by enforcing that the winds only feed the enriched reservoir, we have assumed that metals are efficiently mixed into the ejected outflows and cannot meaningfully penetrate the IGM filaments, which feed the pristine reservoir. Note that material swept up by winds is incorporated into the enriched reservoir after a delay time that is on the order of the halo's dynamical time (and is discussed in more detail in Section~\ref{sec:PopIII-II_transition}). After this time, we do allow for `transfer' of material from the pristine to the enriched reservoir as a proxy for the metal pollution process, as turbulent mixing drives chemical homogenization of the CGM over a timescale $t_{\rm mix}$. 

Before moving to define the remaining terms in these equations, it is instructive to introduce the final components of our coupled system, which dictate how material flows through the system. That is, we need to model the energy evolution of the CGM reservoirs in order to understand the rates at which material will be gained and lost from the various phases. The CGM accumulates energy from cosmological accretion and SN-driven outflows from the galaxy and loses it via turbulent dissipation and as mass escapes from the system. These are separated as the mass reservoirs and satisfy the following evolution equations.
\begin{align} 
\dot{E}_{\rm CGM}^{\rm pri} &= (1-f_{\rm enr})\dot{E}_{\rm acc} - \dot{E}_{\rm diss}^{\rm pri} - \dot{E}_{\rm esc}^{\rm pri}, \\ 
\dot{E}_{\rm CGM}^{\rm enr} &= f_{\rm enr}\dot{E}_{\rm acc} + f_w\dot{E}_{\rm wind} - \dot{E}_{\rm diss}^{\rm enr} - \dot{E}_{\rm esc}^{\rm enr},
\label{eq:E_CGM_enr}
\end{align}
where $\dot{E}_{\rm wind} = \omega_{\rm SN}^{\rm pri}\dot{m}_\star^{\rm pri, \mathcal{D}} + \omega_{\rm SN}^{\rm enr}\dot{m}_\star^{\rm enr, \mathcal{D}}$, with $\omega_{\rm SN}^{\rm (i)}$ being the energy produced by SNe per unit mass of star formation (and so will depend on the IMF), and $f_w = 0.3$ accounts for radiative and other energy losses as the wind travels from the ISM to CGM. We calibrate our choice of $f_{w}$ such that our estimates for the volume filling fraction of metals in the IGM converges to the same ballpark as \cite{liu_when_2020} and \cite{yamaguchi_extent_2023} (see Section~\ref{sec:global_enrichment} for more details) and note that there is mild sensitivity in the level of late-time Pop III star formation to this choice.

As with the associated mass reservoirs, we have assumed that the feedback injection only affects the enriched reservoir and is introduced after a delay time $\mathcal{D}$ (see Section~\ref{sec:PopIII-II_transition}). Energy in excess of the binding energy of the CGM escapes the system over a dynamical time, 
\begin{equation}\label{eq:CGM_energy_escape}
    \dot{E}_{\rm esc}^{\rm (i)} = f_{\rm unb}\frac{E_{\rm CGM}^{\rm (i)} - E_{\rm CGM}^{\rm bin, (i)}}{t_{\rm dyn}},
\end{equation}
where the binding energy is given by $E_{\rm CGM}^{\rm bin, (i)} = 3k_B T_{\rm vir}m_{\rm CGM}^{\rm (i)}/(2\mu m_p)$ and we have taken $f_{\rm unb} \sim 1$. This characterizes the mass outflow rates:
\begin{equation}\label{eq:mass_esc}
    \dot{m}_{\rm esc}^{\rm (i)} = \frac{\dot{E}_{\rm esc}^{\rm (i)}}{E_{\rm CGM}^{\rm (i)}/m_{\rm CGM}^{\rm (i)}}
\end{equation}

Energy is also lost in each of these two phases through a turbulent cascade, which proceeds on a timescale set by the relative scales of turbulent driving $R_{\rm turb}$ and the size of the CGM as a whole $R_{\rm CGM}$ (which we take to be $\sim R_{\rm vir}$). Following the discussion in \cite{pan_modeling_2013}, the CGM will homogenize over a ``self-convolution" timescale that is of order the dynamical time
\begin{equation}\label{eq:t_turb}
    t_{\rm turb}\sim t_{\rm conv} \sim \frac{R_{\rm turb}}{v_{\rm turb}} \sim \mathcal{O}(t_{\rm dyn}).
\end{equation}
Using the results of \cite{pandya_unified_2023} (who appeal to the FIRE galaxy formation simulations to estimate this timescale), we take $R_{\rm turb}\sim R_{\rm vir}$. The turbulent velocity is estimated assuming that the energy in each phase of the CGM is kinetic: 
\begin{equation}\label{eq:v_turb}
    v_{\rm turb}^{\rm (i)} = \sqrt{2E_{\rm CGM}^{\rm (i)}/m_{\rm CGM}^{\rm (i)}}
\end{equation}.

With this in hand, we can return to eqs.~\ref{eq:CGM_pri}-\ref{eq:CGM_Z}. Mass settles out of the two reservoirs on an ``effective" free-fall time 
\begin{equation}\label{eq:m_settle}
\dot{m}_{\rm settle}^{\rm (i)} = \frac{m_{\rm CGM}^{\rm (i)}}{t_{\rm dyn}\sqrt{1+(v_{\rm turb}^{\rm (i)}/v_c)^2}} \equiv\frac{ m_{\rm CGM}^{\rm (i)}}{t_{\rm settle}^{\rm (i)}},
\end{equation}
where turbulent support suppresses this settling.\footnote{\cite{pandya_unified_2023} include an additional cooling timescale in the denominator here, but at the high densities characteristic of high-$z$ galaxies, radiative cooling will proceed on very short timescales and is thus effectively instantaneous. For example, with the central densities estimated as described in Appendix~\ref{app:gas_density} and cooling rates given in \cite{galli_chemistry_1998} (for \molh) and \cite{sutherland_cooling_1993} (for atomic ${\rm H}$), dynamical times are $\mathcal{O}(10\ {\rm Myr})$, while cooling times are $\ll {\rm Myr}$, so the denominator in the analogous version of eq.~\ref{eq:m_settle} in \cite{pandya_unified_2023} is dominated by the dynamical time.}

The turbulent model also allows us to estimate the mixing timescale introduced in eqs.~\ref{eq:CGM_pri}-\ref{eq:CGM_enr}. We use a random walk model for turbulent transport and large-scale mixing will proceed over timescales\footnote{Note that the larger of the turbulent velocity values $v_{\rm turb}^{\rm (i)}$ (eq.~\ref{eq:v_turb}) is chosen to set the mixing time.} 
\begin{equation}\label{eq:t_mix}
t_{\rm mix} \sim f_{\rm mix}\frac{R_{\rm CGM}^2}{R_{\rm turb}v_{\rm turb}},  
\end{equation}
where $f_{\rm mix} = 1$ is a factor that allows us to account for variations from this simple framework. In the limit that $R_{\rm CGM}\sim R_{\rm turb}$, $t_{\rm mix}\sim t_{\rm dyn}$, but if it were smaller, then this term could be adjusted accordingly. Variations in the physical environments of early star-forming halos could modify the mixing time as well. For example, the presence of magnetic fields could confine the extent and modify the velocities of SN-driven outflows, thereby limiting the efficiency and slowing the timescale of metal pollution \citep{van_de_Voort_effect_2021, shah_understanding_2025}. We note, however, that our results are only mildly sensitive to this choice and that our qualitative conclusions do not change for modest variations in $f_{\rm mix}$.

\subsection{The onset of star formation}\label{sec:min_mass}
The formation of the first generations of stars relies on the presence of a sufficient density of molecular hydrogen, H$_2$, to enable efficient cooling in the first (pristine) minihalos. This criterion is often phrased in terms of a critical DM halo mass above which the first stars can form (see e.g., \citealp{tegmark_how_1997, machacek_simulations_2001, kulkarni_critical_2021, schauer_influence_2021, park_population_2021-1, nebrin_starbursts_2023, hegde_self-consistent_2023}). The mass thresholds quoted in such works quantify the response of a halo's ability to form molecular hydrogen to a number of internal and environmental processes. 

Molecular hydrogen buildup is primarily affected by three key processes, which set the critical mass in different epochs. First, a relic relative velocity between DM and baryons imprinted from the epoch of recombination suppresses the ability of a halo to accrete matter from the IGM and cool (discussed in e.g., \citealp{tseliakhovich_relative_2010, dalal_large-scale_2010, naoz_simulations_2012, naoz_simulations_2013, lake_supersonic_2021, williams_supersonic_2023, williams_supersonic_2024, hirano_formation_2025, chen_supersonic_2025}). Once the first generations of stars form and the global star formation rate density (SFRD) begins to rise, a metagalactic background of H$_2$-dissociating, `Lyman-Werner,' photons begins to build up, driving a steep increase in the critical mass (discussed in e.g., \citealp{machacek_simulations_2001, glover_radiative_2003, visbal_high-redshift_2014, kulkarni_critical_2021, schauer_influence_2021}). At the latest times, the remnants from these early stellar populations can contribute to an evolving X-ray background, which contributes to hydrogen ionization and can work both positively and negatively for the star formation process, as the associated heating suppresses accretion but the increased free electron fraction can catalyze H$_2$ formation (discussed in e.g., \citealp{machacek_effects_2003, ricotti_x-ray_2016, park_population_2021-1, park_population_2021-2, hegde_self-consistent_2023}).

In this work, we build on the results of \cite{hegde_self-consistent_2023} and use their calculation of the critical star forming halo mass. We discuss an update to this calculation --- namely, in the calculation of the halo's gas density profile --- in Appendix~\ref{app:gas_density}, but otherwise, our calculation is identical to the aforementioned work.\footnote{In this calculation, we use $\zeta = 0.16$ in the cooling criterion (i.e., we enforce that $t_{\rm cool}\leq t_{\rm ff}=0.16 t_H \implies \zeta = 0.16$), which amounts to estimating a \textit{mean}, or representative, halo mass around which H$_2$ cooling is efficient (see the discussion in Section 5.1 of \citealp{hegde_self-consistent_2023}).} In practice, following the results of \cite{kulkarni_critical_2021} and \cite{feathers_global_2024}, we account for variations in individual halo assembly histories and structure by implementing scatter around this representative mass threshold, drawing from a lognormal distribution with a mean at $m_{\rm crit, H_2}$ and a dispersion of 0.15 dex, though we find that our results are largely insensitive to the exact choice for this dispersion.

Once a halo crosses this star formation threshold, we also enforce that the clouds must be massive enough to overcome the local Jeans mass, i.e., $m_{g}^{\rm pri} > m_J$. In the following sections, once these criteria are met, we implement the ensuing star formation computing average quantities from a Chabrier-like Pop III IMF:\footnote{We note that a previous iteration of the model \citep{hegde_self-consistent_2023} directly sampled the IMF when modelling the star formation process, but we opt not to do so here for simplicity and defer an exploration of the sensitivity to such a choice to future work. We expect that this would introduce small variations in the SFHs of individual halos at a level smaller than many of the other uncertainties outlined here.}
\begin{equation}\label{eq:IMF}
    \frac{dN}{dm}\propto m^{-\alpha}\exp\Bigg[-\bigg(\frac{m_{\rm char}}{m}\bigg)^\beta\Bigg],
\end{equation}
which is a free parameter of our model (Table~\ref{tab:constants}).

\subsection{Feedback and the star formation efficiency}\label{sec:rad_feedback}
Much like the Pop III IMF, theoretical estimates for the star formation efficiency in pristine gas clouds are fairly uncertain. As such, we motivate our choices with a simple physical argument inspired by the simulations of \cite{grudic_when_2018}.

In particular, the densities and metallicities of early star forming environments suggest that it may be important to consider other feedback mechanisms beyond just SN energy or momentum injection to determine the star formation efficiency. For example, in the dust-poor, high density environments characteristic of the first star-forming clouds, Ly$\alpha$ photons produced by the first stars may scatter many times before escaping the cloud, thereby imparting substantial momentum into the surrounding gas and suppressing further star formation \citep{nebrin_Lya_2025}. Such processes, which would begin effectively instantaneously once the first stars form, could thereby act to limit the integrated star formation efficiency \textit{before SN feedback can set in}. 

To account for this effect, we follow the arguments in Section 2.2 of \cite{grudic_when_2018} (also \citealp{fall_stellar_2010}) and balance the forces of gravity and feedback to identify the point at which stellar feedback would be strong enough to break apart the cloud and quench star formation. The force of gravity binding the gas to the cloud is
\begin{equation}\label{eq:grav_force_cloud}
    F_g = \frac{G(m_g + m_\star)m_g}{r_{\rm cloud}^2}.
\end{equation}
As stars form, radiative feedback sets in, driving a force
\begin{equation}\label{eq:rad_force}
    F_{\rm rad} = \bigg\langle \frac{P_{\rm rad}}{m_\star}\bigg\rangle \dot{m}_\star,
\end{equation}
where $\langle P_{\rm rad}/m_\star\rangle$ is the IMF-averaged momentum injected per unit mass of star formation given by
\begin{align}\label{eq:rad_momentum}
\begin{split}
    \bigg\langle \frac{P_{\rm rad}}{m_\star}\bigg\rangle &= \frac{\overline{E}_{\rm ion}\langle N_{\rm ion}/m_\star\rangle + M_{\rm F}E_{\alpha}\langle N_{\rm \alpha}/m_\star\rangle}{c}\big(1-e^{-\tau_{\rm HI}}\big) \\
    &= \frac{\overline{E}_{\rm ion} + \frac{2}{3} M_{\rm F}E_{\alpha}}{c}\bigg\langle\frac{N_{\rm ion}}{m_\star}\bigg\rangle\big(1-e^{-\tau_{\rm HI}}\big),
\end{split}
\end{align}
for which the escape fraction of ionizing photons is given by $f_{\rm esc, ion} = e^{-\tau_{\rm HI}}$. The IMF-averaged quantities (denoted with $\langle \cdot \rangle$) are computed using the Pop III stellar tables from \cite{schaerer_properties_2002}; e.g.,
\begin{equation}\label{eq:IMF_avg_photonCounts}
    \bigg\langle\frac{N_{i}}{m_\star}\bigg\rangle = \frac{\int_{m_{\rm min}}^{m_{\rm max}}Q_i(m)t_{\rm life}(m)\phi(m)dm}{\int_{m_{\rm min}}^{m_{\rm max}}m\phi(m)dm}
\end{equation}
is the mean number of ionizing/dissociating photons produced per unit stellar mass (for a species $i = {\rm H,\ HeI,\ HeII,\ H_2}$), accounting for the different stellar lifetimes $t_{\rm life}(m)$. The factor of 2/3 comes from the fact that roughly 2/3 of recombinations produce a Ly$\alpha$ photon, and the Ly$\alpha$ force multiplier $M_{\rm F}(N_{\rm HI}, T, Z, z)$ is given by the fits found in \cite{nebrin_Lya_2025}. \footnote{In this expression we set $T\sim T_{\rm vir}$ and $N_{\rm HI}\sim n_{\rm core}\times r_{\rm ISM}\sim n_{\rm core}\times 0.1 r_{\rm vir}$, where $n_{\rm core}$ is estimated following the calculation outlined in Appendix~\ref{app:gas_density}.} 

Defining $\epsilon \equiv m_\star / (m_g + m_\star)$ and equating eqs.~\ref{eq:grav_force_cloud} and \ref{eq:rad_force} (and approximating $\dot{m}_\star \sim m_\star / t_{\rm burst} = m_\star / \langle t_{\rm life}\rangle$), we find that 
\begin{equation}\label{eq:sfe_max}
    \epsilon_{\rm max} = \frac{m_\star + m_g}{m_{\rm crit} + m_\star + m_g},
\end{equation}
where $m_{\rm crit} = \alpha_{\rm rad}\langle P_{\rm rad}/m_\star\rangle r_{\rm cloud}^2/\langle t_{\rm life} \rangle G$, $r_{\rm cloud} \sim r_{\rm ISM} = 0.1 r_{\rm vir}$, and $\alpha_{\rm rad}$ accounts for the uncertainties folded into this calculation (e.g., the true size of the star-forming clouds and the details of the radiative transfer within them). For $\alpha_{\rm rad}\sim 0.25$, we find comparable results to other semi-analytic calculations (which assume $\epsilon^{\rm III}\sim 0.001$; e.g., \citealp{visbal_self-consistent_2020, ventura_semi_2025}). We explore the variations induced by this choice in Section~\ref{sec:early_SFRD}. This sets a maximum on the integrated star formation efficiency that roughly evolves from $\mathcal{O}(10^{-2})$ at $z\gtrsim 30$ to $\mathcal{O}(10^{-4}-10^{-3})$ at $z\sim 10-20$. With this in hand, we set the star formation efficiency per free fall time $\epsilon_{\rm ff}^{\rm III} = 1$, but limit the stellar mass that can be formed in any burst by the maximum integrated efficiency computed with eq.~\ref{eq:sfe_max}. We note that the star formation criteria and feedback effects estimated throughout apply to reservoir-averaged quantities, rather than individual clumps, and we defer the additional sophistication of a small-scale star formation model to future work.

\subsection{The Pop III-II transition}\label{sec:PopIII-II_transition}
Once the first burst of stars forms, we prescribe that their SNe feedback is injected into the ISM gas after a time set by the mean lifetime of the stars in the chosen Pop III IMF, $\langle t_{\rm life}\rangle$, at which point gas is driven into the enriched CGM reservoir. However, simulations demonstrate that this gas is not instantaneously reaccreted --- instead, there is a characteristic `recycling time' over which the gas is incorporated into the turbulent CGM and becomes eligible for reaccretion (see e.g., \citealp{chiaki_metal-poor_2018, anglesalcazar_cosmic_2017, hopkins_what_2023}). These simulations suggest that gas does not begin to recollapse until a delay on the order of the halo's dynamical time has passed. Motivated by this, we implement a delay $t_{\rm incorp} \sim t_{\rm dyn}$ before the outflowing wind mass is incorporated into the enriched reservoir.\footnote{Numerically, this amounts to placing the second and third terms of eq.~\ref{eq:CGM_enr} into a temporary `wind mass' reservoir until a dynamical time has passed, after which it is incorporated into $m_{\rm CGM}^{\rm enr}$ over a dynamical time.} On an individual system level, this delay time most significantly modulates the early star formation history of the halo, though it is present throughout the galaxy's history. At a population scale (see Section~\ref{sec:pop_quantities}), this determines the point at which we transition from Pop III to II dominance in the global star formation rate density. 

\begin{figure*}
    \centering
    \includegraphics[width=0.9\linewidth]{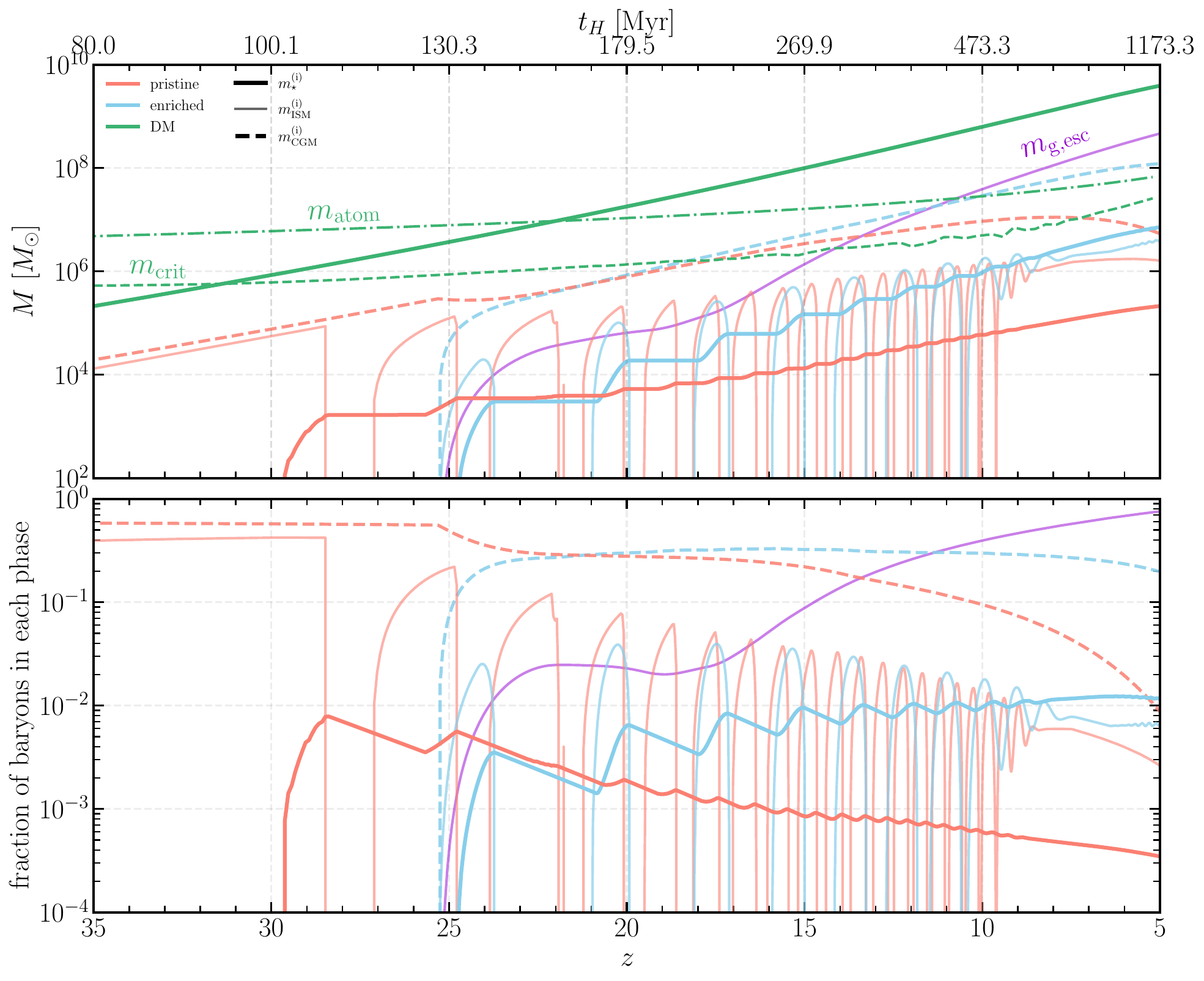}
    \caption{\textbf{In the absence of environmental effects, inefficient mixing of metals ejected by stellar feedback events can result in a high level of Pop III star formation out to $z\sim 5$.} Evolution of a $\sim 10^{9.5}\ M_\odot$ halo from $z=35$ to $z=5$. (top) The different mass reservoirs characterizing the halo are indicated with colored lines, with pristine reservoirs identified by red, enriched by blue, and DM by green. For a given (red or blue) color, different line styles denote the different reservoirs --- dashed are the CGM gas reservoirs, thin solid are the ISM, and thick solid are the cumulative stellar mass formed over the galaxy's history. The cumulative mass that has escaped the system due to stellar feedback is shown with a purple solid line. For reference, the critical halo mass for Pop III star formation (an environmental quantity) is shown with a green dashed line and the atomic cooling threshold is indicated with a green dot-dashed line. (bottom) The corresponding fractions of baryonic mass for each of the curves in the upper panel; i.e., $X_i = m_i/f_bm_h$.}
    \label{fig:SFH_ex}
\end{figure*}

\subsection{Shutting off later star formation}\label{sec:LW_feedback}
Once stars form and the system begins to evolve, repeating cycles of Pop II and III star formation proceed, according to the delayed feedback behavior described above. One complexity that has thus far been ignored is the mutual feedback between the Pop II and III stellar populations. To get a handle on this effect, we ignore the mechanical feedback introduced by SNe (essentially assuming that spatial segregation of star formation is sufficient to keep such feedback events independent), but introduce another source of radiative feedback, namely in the form of molecular hydrogen-dissociating UV photons in the Lyman-Werner (LW) bands (11.2-13.6 eV). As discussed in Section~\ref{sec:min_mass}, star formation in pristine gas relies on \molh\ cooling to proceed, so the ability for a cloud to shed its thermal energy and undergo runaway collapse is crucially tied to the density of \molh. While the minimum mass criterion, at least as described in \cite{hegde_self-consistent_2023}, is no longer appropriate to characterize the onset of later generations of star formation after a halo's first star formation event, the physics incorporated into that calculation are still relevant. In particular, feedback from an internal LW field will likely suppress star formation in the pristine reservoir. 

To estimate the strength of this effect, we compute the local LW intensity produced by the current stellar populations present in our galaxy using a similar procedure to that outlined in Section~\ref{sec:rad_feedback}. That is, for Pop III stars, the LW luminosity is given by $\langle L_{\rm LW}^{\rm III}\rangle = \langle N_{\rm LW}/m_\star\rangle \overline{E}_{\rm LW}\dot{m}_\star$. For Pop II stars --- where we do not have an explicit stellar table --- we use a similar expression, taking the number of LW photons per stellar baryon to be $N_{\rm LW}^{\rm II}\sim 9690$ so $L_{\rm LW}^{\rm II} = N_{\rm LW}^{\rm II} \overline{E}_{\rm LW}\dot{m}_\star/m_p$ (see e.g., \citealp{leitherer_starburst99_1999, mebane_persistence_2018}). With these in hand, the local mean LW intensity can be computed as
\begin{equation}
    J_{\rm LW}^{\rm local} = \frac{1}{4\pi}\int I_{\nu}d\Omega = \frac{dE}{dtd\nu dA} \approx \frac{\langle L_{\rm LW}^{\rm III}\rangle + L_{\rm LW}^{\rm II}}{\Delta \nu_{\rm LW}r_{\rm ISM}^2} 
\end{equation}
For a $5\times 10^8\ M_\odot$ halo at $z=15$, assuming that all the LW photons escape their HII regions, we find that $J_{\rm LW, 21}^{\rm local, II} \sim 5\times 10^{7}\ \dot{m}_{\star}^{\rm II}$, where the intensity is now expressed in units of $J_{21} = 10^{-21}\ {\rm erg\ s^{-1}\ cm^{-2}\ Hz^{-1}\ sr^{-1}}$ and the SFR is in units of $M_\odot\ {\rm yr^{-1}}$. From \cite{klessen_first_2023}, given a gas density $n$ and self-shielding factor $f_{\rm sh}$ (the factor by which the dissociation rate is suppressed by self-shielding in dense gas), the critical intensity for efficient dissociation (from balancing the dissociation and recombination/\molh\ formation times) is $J_{\rm crit, 21}\sim 10^{-4} f_{\rm sh}n$, which is roughly $\mathcal{O}(10^{-10})$ for the aforementioned $5\times 10^8\ M_\odot$ halo. As a result, even if the Pop II-generated local LW field were heavily suppressed (e.g. if $f_{\rm esc}$ were low, there was significant self-shielding in the pristine clouds, etc.) a fairly low level of Pop II star formation can still effectively shut off H$_2$ cooling, as this would easily raise $J_{\rm LW}$ well beyond that threshold. Therefore, to implement this effect, once Pop II star formation begins, we `turn off' Pop III star formation in halos \textit{until they have crossed the atomic cooling threshold and can appeal to Ly$\alpha$ cooling to drive star formation in their pristine clouds.}\footnote{In principle, this cooling process could result in a different IMF compared to the \molh cooling (i.e., the difference between `Pop III.1' and `III.2'), but, given the vast uncertainties in the IMF at low metallicity and high-redshift (e.g., \citealp{hirano_one_2014, hirano_primordial_2015, stacy_building_2016, jaura_trapping_2022, chon_impact_2024}), we ignore that complexity and assume that all star formation in pristine gas follows the same IMF (eq.~\ref{eq:IMF}), though we plan to explore that sensitivity further in future work.}

\subsection{An example galaxy history}\label{sec:internal_summary}
In Figure~\ref{fig:SFH_ex}, we show an implementation of the model --- with just the internal processes (i.e., the aforementioned components described in this section) --- for an example galaxy history in a $m_h(z=5)\sim 10^{9.5}M_\odot$ halo. We can understand the response of the galaxy to the various model components by walking through this representative history. First, the halo grows smoothly and the CGM and ISM pristine gas reservoirs track that accordingly. At $z\sim 31$, this halo crosses the critical mass threshold (Section~\ref{sec:min_mass}) and is in principle eligible to form stars. However, the additional Jeans criterion (Section~\ref{sec:min_mass}) for star formation induces an additional delay of a few Myr until star formation commences at $z\sim 28$. 

At this point, a burst of Pop III, metal-free stars forms, with an integrated efficiency $\mathcal{O}(10^{-3})$ (Section~\ref{sec:rad_feedback}), and star formation proceeds until the first SNe explode roughly 5 Myr later. At the low masses characteristic of these first star forming halos, the binding energy of the gas is sufficiently low and the more energetic SN feedback of the first metal free stars (Section~\ref{sec:two_phase_ISM}) is able to efficiently evacuate the ISM reservoir and begin populating the enriched CGM. However, there is a recycling time delay $\mathcal{O}(t_{\rm dyn})$ for these winds to be eligible to settle into the ISM and promote the next generation of enriched star formation (Section~\ref{sec:PopIII-II_transition}). 

During this time, pristine gas continues to fall into the ISM and another burst of Pop III stars forms around $z\sim 25$. Shortly thereafter, the first Pop II burst begins and the internal radiation field produced by these stars temporarily suppresses further Pop III star formation (Section~\ref{sec:LW_feedback}). At $z\sim 23$, the halo crosses the atomic cooling threshold and star formation is permitted to begin again in the pristine gas, leading to cycles of star formation out of both the pristine and enriched ISM reservoirs, punctuated by feedback events that temporarily quench the galaxy. Coincident with this, the repeated ejection of metals from the ISM and injection into the CGM prompts the mixing process (eq.~\ref{eq:t_mix}), which leads the enriched CGM reservoir to steadily grow as the pristine reservoir is polluted. Ultimately, as the system approaches equilibrium at the latest times, this results in \textit{ongoing star formation in both reservoirs}, with a cumulative stellar mass of $\sim 10^5M_\odot$ in the pristine reservoir and $\sim 10^{6} M_\odot$ in the enriched, an $\mathcal{O}(10\%)$ ratio between the two stellar populations, essentially just set by the ratio of the integrated star formation efficiencies that result from this framework.

\section{External modifications}\label{sec:external_modifications}
While the framework introduced in the preceding section is sufficient for a galaxy in isolation, which is largely the case for the early stages of the galaxy's evolution, the late-time behavior will be highly sensitive to more global processes that affect the galaxy's enrichment (such as the growth of enriched wind bubbles in the IGM) and star formation (such as reionization heating of the IGM).

\subsection{Enrichment of the IGM}\label{sec:global_enrichment}
As the first star formation and feedback episodes proceed, they will eject metals into the circum- and intergalactic media, beginning the process of local and global IGM enrichment. First, we estimate the effect of the former of these processes: the pollution of metals into the accretion filaments feeding a galaxy.

Once stellar feedback events are able to drive mass out of a halo, winds carrying enriched material can in principle begin to enrich infalling, previously-pristine gas. To quantify the strength of this effect, we appeal to the timescale over which local winds passing by a filament drive the formation of fluid instabilities between the media and thus could begin to efficiently mix their material with the inflowing gas. That is, once a halo begins ejecting material beyond its virial radius, we estimate the Kelvin-Helmholtz timescale in a filament as
\begin{equation}
    \label{eq:KH_time}
    t_{\rm KH}(m_h) = \frac{\lambda_J}{v_{\rm wind}}\frac{2+\delta}{\sqrt{1+\delta}},
\end{equation}
where $\delta\sim 40$ is the characteristic density contrast of cold IGM filaments relative to the mean density of the IGM (see e.g., \citealp{mandelker_cold_2018, lu_structure_2024}), $\lambda_J$ is the Jeans length in the IGM (which evolves as the IGM is heated) and is taken to represent the filament size \citep{schaye_thermal_2000}, and the wind velocity is estimated as the larger of the escape velocity and the velocity of the infalling material, $v_{\rm wind} = \max\big(v_{\rm esc}, v_{\rm infall}\big)$. This characterizes the timescale for pollution of an inflowing gas parcel travelling along a filament. If the associated infall time for this parcel is long compared to its KH time, then instabilities can grow during infall and mixing can efficiently proceed, so we then assume that this gas would be enriched. In the opposite limit, gas falls into a halo more rapidly than instabilities can grow, so pollution from local enrichment is inefficient. In practice, we quantify the level of self-enrichment from a halo's early star formation episodes by the following:
\begin{equation}
    \label{eq:self-enrichment}
    f_{\rm enr, local} = \frac{t_{\rm infall}}{t_{\rm instability}} = \frac{t_{\rm ff}(3\times r_{\rm vir}, z)}{t_{\rm KH}(m_h)},
\end{equation}
where we have taken the infall time to be the free-fall time evaluated at 3 times the halo's virial radius. 

At the same time, in aggregate, winds from a population of star-forming halos will begin the gradual enrichment of the bulk IGM. While the growth of these bubbles has been studied in more detail in a number of previous works (see e.g., \citealp{tegmark_late_1993, madau_early_2001, furlanetto_metal_2003, liu_when_2020, yamaguchi_extent_2023, ventura_role_2023}), here we are primarily concerned with estimating the probability that a halo is forming in an enriched bubble. To this end, we follow the simplifications outlined in \cite{furlanetto_is_2005} (which were shown to provide a good approximation to the more detailed calculations of \citealp{furlanetto_metal_2003}) to estimate the volume filling fraction of metals in the IGM.

\begin{figure}
    \centering
    \includegraphics[width=0.9\linewidth]{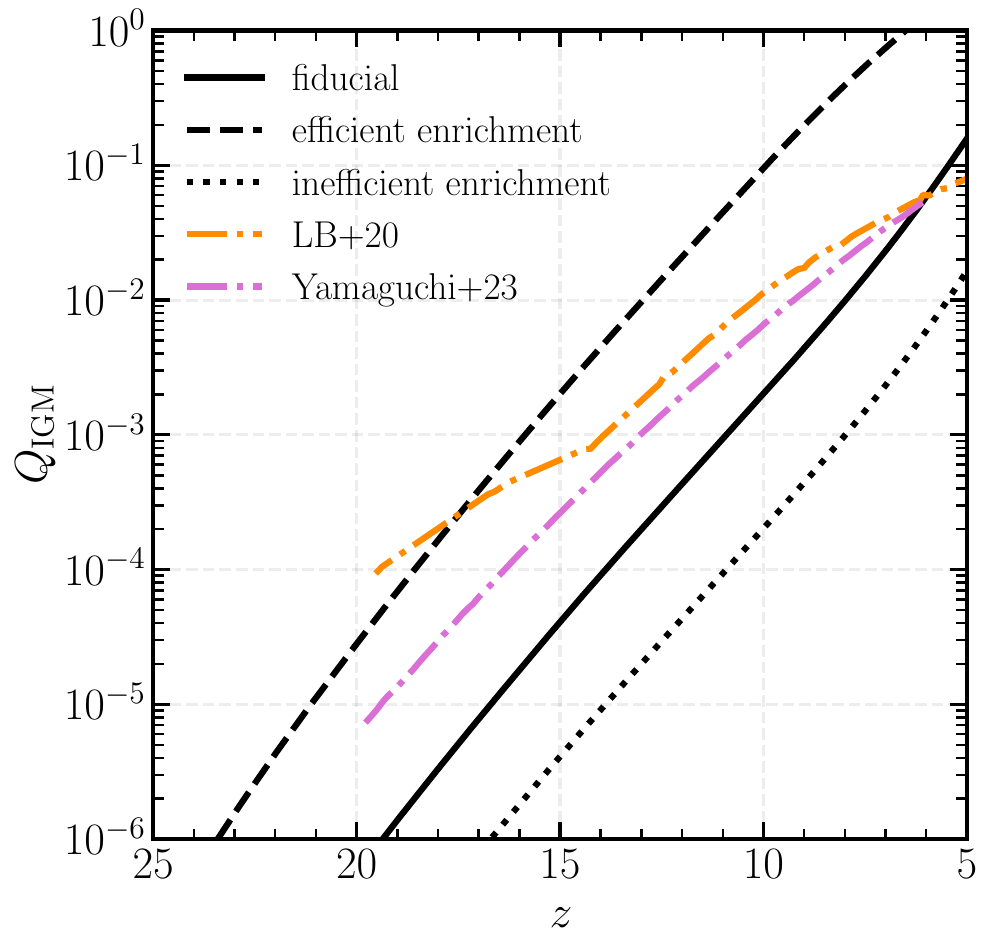}
    \caption{The estimated evolution of $Q_{\rm IGM}$ following the framework discussed in Section~\ref{sec:global_enrichment} for three choices of the efficiency value for the energy carried by stellar winds (1, 50, and 0.1 $\times K_w$ for the solid, dashed, and dotted curves, respectively). For reference, we also show the model estimates from \cite{liu_when_2020} and \cite{yamaguchi_extent_2023} as orange and purple dot-dashed curves, respectively.}
    \label{fig:QIGM_evolution}
\end{figure}

In particular, our galactic wind model follows the approximate \cite{sedov_similarity_1959} wind solution, which describes an energy-conserving explosion expanding into a constant density medium. This solution predicts that the wind radius scales as $r_{\rm bubble}^{\rm metal} \approx 1.17(E_{\rm SN}t^2/\rho)^{1/5}$, where $\rho$ is the ambient density, $t$ is the time from explosion, and $E_{\rm SN}$ is the total energy produced in SNe escaping from a given DM halo (which is a natural product of our galaxy model; eq.~\ref{eq:CGM_energy_escape}). This solution allows us to express the mass loading factor of winds $\eta_{\rm metal} \equiv m_{\rm enr}/m_h$ as
\begin{equation}\label{eq:eta_metal}
    \eta^{\rm metal}= K_w\frac{m_{\rm enr}}{m_h} = \frac{4\pi K_w}{3}\frac{\big[r_{\rm bubble}^{\rm metal}(E_{\rm esc})\big]^3\rho_{\rm IGM}}{m_h} ,
\end{equation}
where $K_w$ is a normalization constant that accounts for the simplifications made in this estimate, $E_{\rm esc}$ is the energy escaping the halo (eq.~\ref{eq:CGM_energy_escape}), and $\rho_{\rm IGM}$ is the mean density of the IGM. Motivated by the results of \cite{furlanetto_is_2005} (who calibrate this model to their simulations of expanding wind bubbles), we choose $K_w\sim (1/27) f_{w, \rm FL05}^{3/5}$, where $f_{w, \rm FL05} = \min(1, t_{\rm comp}/t_H)$ (with $t_{\rm comp} = 1.2\times 10^8\ [10/(1+z)]^4{\rm\ yr}$; \citealp{furlanetto_minimalist_2017}) accounts for radiative losses as the winds progress. 

As a proxy for estimating the relative rates with which gas is accreted into the enriched and pristine reservoirs on large scales, we estimate the probability that any given halo sits in an enriched bubble, which would dominate in the limit that \textit{mixing of metals into accretion filaments proceeds inefficiently (i.e., $t_{\rm KH}$ is long in eqs.~\ref{eq:KH_time}-\ref{eq:self-enrichment}), and thus tracks the global enrichment of the IGM}. With the mass loading factor, we can estimate the fraction of the IGM filled with metals:
\begin{equation}
    Q_{\rm enr}^{\rm IGM} = \int dm_h \frac{m_h}{\overline{\rho}_b}\eta_{\rm metal}(m_h)n(m_h)
\end{equation}
The results of this calculation for a few different choices of the normalization of eq.~\ref{eq:eta_metal} are shown in Figure~\ref{fig:QIGM_evolution}, from which it can be seen that our fiducial model (discussed in more detail in Section~\ref{sec:pop_quantities}) results in an IGM that is $\sim 10-20\%$ enriched by $z\sim 5$. We note that our results here are broadly consistent with the other model-based estimates of this quantity presented in \cite{liu_when_2020} and \cite{yamaguchi_extent_2023}.

Because halo formation is a clustered process, we then estimate the probability that a halo lies within the wind radius of a nearby halo using the galaxy two-point correlation function $\xi_{\rm gg}\sim b_1b_2\xi_{\delta\delta}$, where $b_1$ and $b_2$ are the associated host halo biases. To this end, the probability that a halo lies within a radius $r_w$ of another halo is
\begin{equation}\label{eq:clustering}
    p_{\rm near}(m_h) = Q_{\rm enr}^{\rm IGM}\big[1 + b(m_h)\overline{b}_{\rm metal}\xi_{\delta\delta}(r_w)\big],
\end{equation}
where we again use the Sedov solution to estimate $r_w$ as is done in \cite{furlanetto_minimalist_2017}, motivated by the simulations of \cite{furlanetto_metal_2003}. The mean bias of enriched bubbles is
\begin{equation}
    \overline{b}_{\rm metal} = \frac{\int dm_h m_h \eta_{\rm metal}(m_h)b(m_h)n(m_h)}{\int dm_h m_h n(m_h)}.
\end{equation}
This clustering-based treatment of the enrichment probability enhances the local value by a factor of roughly 2-4 and is only mildly dependent on halo mass. Assuming that the wind hosts are randomly distributed, the probability that a halo is in an enriched region is
\begin{equation}
    \label{eq:f_enr}
    f_{\rm enr, global}(m_h) = 1 - \exp\big[-p_{\rm near}(m_h)\big].
\end{equation}

For simplicity, we assume that the overall level of enrichment is determined by the more effective of these two channels (i.e., we choose the larger of the self-enrichment and global enrichment effects, eqs.~\ref{eq:self-enrichment} and ~\ref{eq:f_enr}). The parameter that characterizes how inflowing material is distributed between the two reservoirs in eqs.~\ref{eq:CGM_pri} and \ref{eq:CGM_enr} is then:
\begin{equation}
    f_{\rm enr} = \max\big(f_{\rm enr, local}, f_{\rm enr, global}\big)
\end{equation}
In practice, we find that self-enrichment dominates, as the IGM is only slightly enriched globally at late times (as can be seen in Figure~\ref{fig:QIGM_evolution}). We note that our prescription is moderately sensitive to the choice of how to model local and global IGM enrichment and emphasize that future numerical work is needed to better understand these processes.

\subsection{Reionization feedback}\label{ssec:reion_feedback}
As metal wind bubbles grow and propagate through the IGM, so too do ionization fronts, as the process of reionization commences with the birth of the first stars. While it takes time for these ionized regions to significantly affect the global state of the IGM, they will have important implications for the late stages of halo growth ($z\lesssim 15$), especially in the lowest mass systems. That is, as the environment around a halo is ionized and heated, later accretion onto the halo is suppressed. A number of studies have explored this effect (see e.g., \citealp{shapiro_reionization_1994, gnedin_probing_1998, gnedin_effect_2000, hoeft_dwarf_2006, okamoto_mass_2008, sobacchi_how_2013, sobacchi_depletion_2013, wu_simulating_2019, katz_how_2020}) and a physical description of the process is presented in detail in \cite{noh_physical_2014}. 

In practice, suppression of accretion onto a halo will reduce the gas fraction, which appears in eqs.~\ref{eq:CGM_pri} and \ref{eq:CGM_enr} through $\dot{m}_{\rm g, acc} \sim f_g f_b\dot{m}_h$. The simplest parameterization of this effect is given in \cite{sobacchi_depletion_2013}, where the critical halo accretion threshold is described by the following:
\begin{equation}\label{eq:Mcrit_acc}
    m_{\rm crit}^{\rm acc} = 10^{9.45}\ M_\odot\ J_{\rm UV, 21}^{0.17}\bigg(\frac{1+z}{10}\bigg)^{-2.1}\Bigg[1-\bigg(\frac{1+z}{1+z_{\rm IN}}\bigg)^2\Bigg]^{2.5},
\end{equation}
where $z_{\rm IN}$ is the redshift at which the halo is first exposed to a UV background of intensity $J_{\rm UV,21}$ (in units of $J_{21}$). We then model the gas fraction suppression with a soft step function about that mean mass, using
\begin{equation}\label{eq:fgas}
    f_g(m_h) = \frac{1}{2}\Bigg[1 + \tanh\bigg(\frac{\log_{10}m_h - \log_{10}m_{\rm crit}^{\rm acc}}{\Delta\log m_{\rm acc}}\bigg)\Bigg],
\end{equation}
with $\Delta\log m_{\rm acc} = 0.5$, which approximately reproduces the results of \cite{noh_physical_2014}.

\begin{figure*}
    \centering
    \includegraphics[width=1\linewidth]{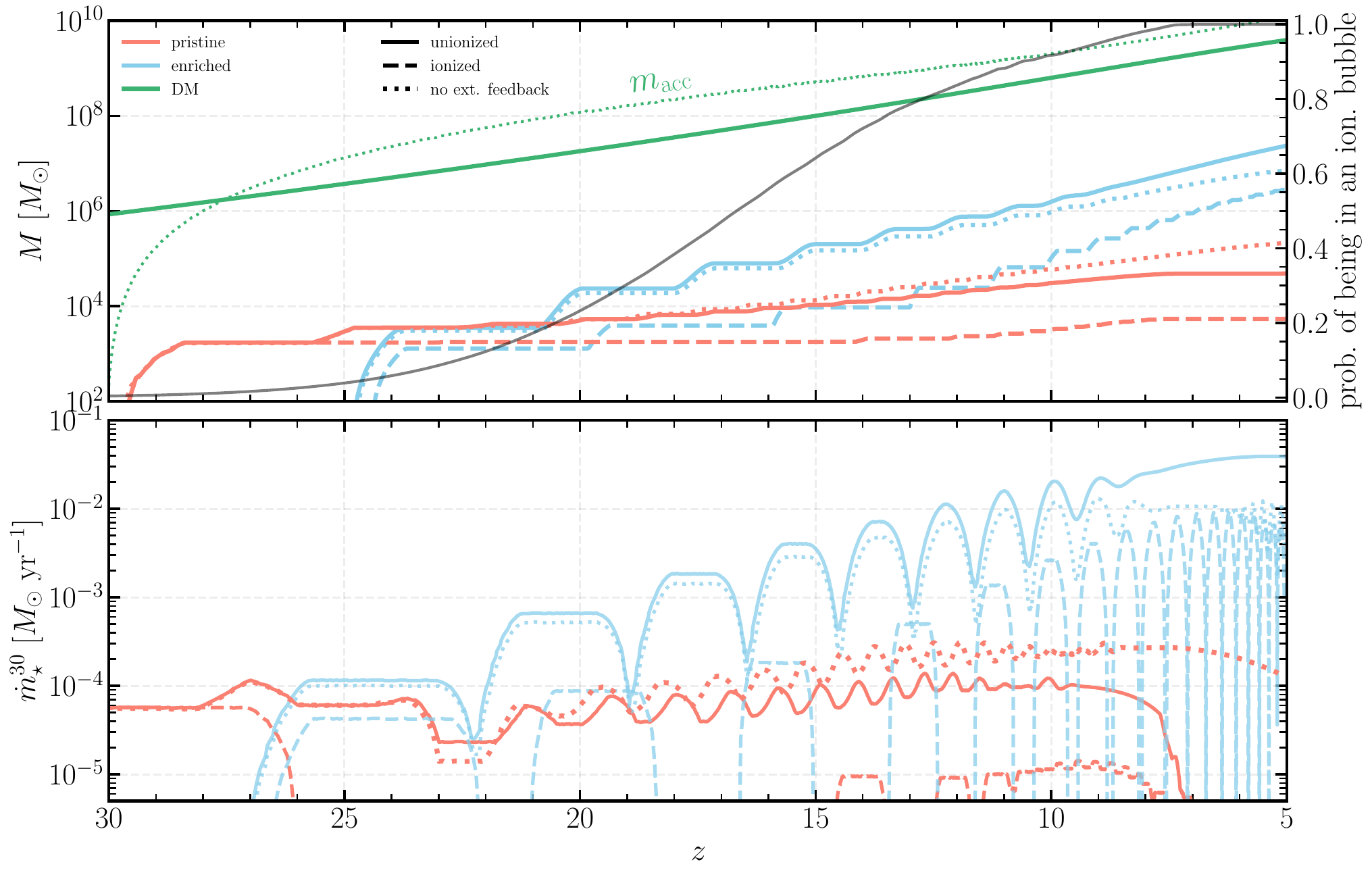}
    \caption{\textbf{Reionization feedback and IGM enrichment can significantly influence the late time evolution of low-mass halos.} \textit{(upper)} Evolution of the same $\sim 10^{9.5}\ M_\odot$ halo from Figure~\ref{fig:SFH_ex}, now with external feedback processes included. We additionally display the accretion threshold induced by reionization heating of the IGM (eq.~\ref{eq:Mcrit_acc}) as a green dotted line. The cumulative stellar mass if the halo were in an (un)ionized region is shown with a dashed (solid) line, with the associated probability denoted by the black line (corresponding to the right vertical axis). The corresponding stellar mass curves in the absence of external feedback (i.e., the same as Figure~\ref{fig:SFH_ex}) are shown as colored dotted curves. \textit{(lower)} Star formation rate smoothed over a 30 Myr window in the pristine and enriched reservoirs for the same ionized and unionized cases, with the pristine SFR from Figure~\ref{fig:SFH_ex} dotted again.}
    \label{fig:SFH_reionization_feedback}
\end{figure*}

In order to estimate this critical accretion threshold, we need to evolve the global UV background. To do this, we solve the cosmological radiative transfer equation and integrate the following:
\begin{equation}\label{eq:J_UV}
    J_{\rm UV}(z) = \frac{c}{4\pi}\int_z^\infty dz' \frac{d\ell}{dz}\frac{(1+z)^3}{(1+z')^3}\epsilon_{\rm UV}(z')e^{-\tau},
\end{equation}
where $d\ell/dz = c/[H(z)(1+z)]$ is the proper line element and the optical depth is given by $\tau = \int_z^{z'} (d\ell / dz'') dz''[n_{\rm HI}\sigma_{\rm HI}(\nu'') + n_{\rm HeI}\sigma_{\rm HeI}(\nu'')]$.\footnote{For the cross sections we use the fits of \citet{verner_atomic_1996}.} The proper emissivity is given by 
\begin{equation}\label{eq:uv_emissivity}
    \epsilon_{\rm UV}(z) = \Bigg[\frac{\rho_{\rm SFR}^{\rm II}(z)}{\kappa_{\rm UV}} + \frac{\rho_{\rm SFR}^{\rm III}(z)}{\kappa_{\rm UV}^{\rm III}}\Bigg](1+z)^3,
\end{equation}
where $\kappa_{\rm UV} = 1.15\times 10^{-28}\ M_\odot {\rm \ yr^{-1}\ (erg\ s^{-1}\ Hz^{-1})^{-1}}$ is a standard conversion factor from SFR to UV luminosity for a Salpeter IMF \citep{madau_cosmic_2014} and $\kappa_{\rm UV}^{\rm III}$ is estimated by integrating over the IMF as before (eq.~\ref{eq:IMF_avg_photonCounts}), resulting in $\kappa_{\rm UV}^{\rm III}\sim 6.18\times 10^{-29}\ M_\odot {\rm \ yr^{-1}\ (erg\ s^{-1}\ Hz^{-1})^{-1}}$ for our fiducial IMF.

To account for the inhomogeneity of the reionization process, we implement this threshold by tracking two copies of every halo in our model: one that grows in an ionized region (with the strength of the reionization suppression driven by the global UV background) and one that evolves without the reionization feedback (and so would be growing in a neutral island). When computing a global quantity like the star formation rate density (SFRD), we then weight these two histories by the probability that each halo lies in an ionized region, which steadily increases as ionized bubbles grow and the IGM becomes ionized.

This then requires an estimate of the global ionized fraction. To compute this, we follow \cite{furlanetto_minimalist_2017} and \cite{munoz_reionization_2024} and solve the `reionization equation' \cite{madau_radiative_1999}
\begin{equation}
    \label{eq:reionization}
    \frac{dx_{\rm HII}}{dt} = \frac{f_{\rm esc}\dot{n}_{\rm ion}}{n_{\rm H}} - \frac{x_{\rm HII}}{t_{\rm rec}},
\end{equation}
where $t_{\rm rec} = [C\alpha_B(1+x_{\rm He})n_H]^{-1}$ is the recombination time for a He fraction $x_{\rm He} \approx Y_{\rm He}/[4(1-Y_{\rm He})]$ and the comoving number density of ionizing photons produced per unit time is given by
\begin{align}\label{eq:nion_III}
    \dot{n}_{\rm ion}^{\rm III}  &= \int dm_h \frac{dn}{dm_h}\bigg\langle \frac{N_{\rm ion}}{m_\star}\bigg\rangle \dot{m}_\star^{\rm III}  \\
    \dot{n}_{\rm ion}^{\rm II}  &= \int dm_h \frac{dn}{dm_h} \frac{\dot{m}_\star^{\rm II}}{\kappa_{\rm UV}}\xi_{\rm ion} \label{eq:nion_II},
\end{align}
where the ionizing efficiency, $\xi_{\rm ion} = 10^{25.29}\ {\rm Hz\ erg^{-1}}$ is taken from the latest JWST spectroscopic constraints \citep{pahl_spectroscopic_2025} and $\kappa_{\rm UV}$ is defined as above. For our fiducial Pop III Chabrier-like IMF, we have $\langle N_{\rm ion}/m_\star \rangle\sim 7.71\times 10^{61}\ M_\odot^{-1}$, when averaged over the the stellar lifetime (see eq.~\ref{eq:IMF_avg_photonCounts}).

For simplicity, we assume a mass-independent escape fraction $f_{\rm esc}^{\rm pri} = f_{\rm esc}^{\rm enr}= 0.1$, set $C\approx 3$, and initialize the ionized fraction at early times to be $x_{\rm HII}=2\times 10^{-4}$ \citep{cosmorec}. We integrate eq.~\ref{eq:reionization} alongside the galaxy evolution equations to track the evolution of the global ionized fraction. With $x_{\rm HII}$ in hand, the probability that a halo sits in an ionized region is modified by the halo clustering (as was done in the enrichment calculation, see eq.~\ref{eq:clustering}), where here we estimate the size of a halo's local ionized bubble by counting the total number of ionizing photons it has produced (which is derived from the Pop III and II star formation rates) and comparing that to the density of hydrogen in the IGM:
\begin{equation}
    r_{\rm bubble}^{\rm ionized} = \bigg(\frac{3N_{\rm tot}^{\rm ion}}{4\pi n_{\rm IGM}}\bigg)^{1/3}.
\end{equation}
Similarly to the case of local IGM enrichment, the clustering treatment modifies the ionization probability by a factor of 2-5, but in this case more steeply varies with halo mass.

\begin{figure*}
    \centering
    \includegraphics[width=0.9\linewidth]{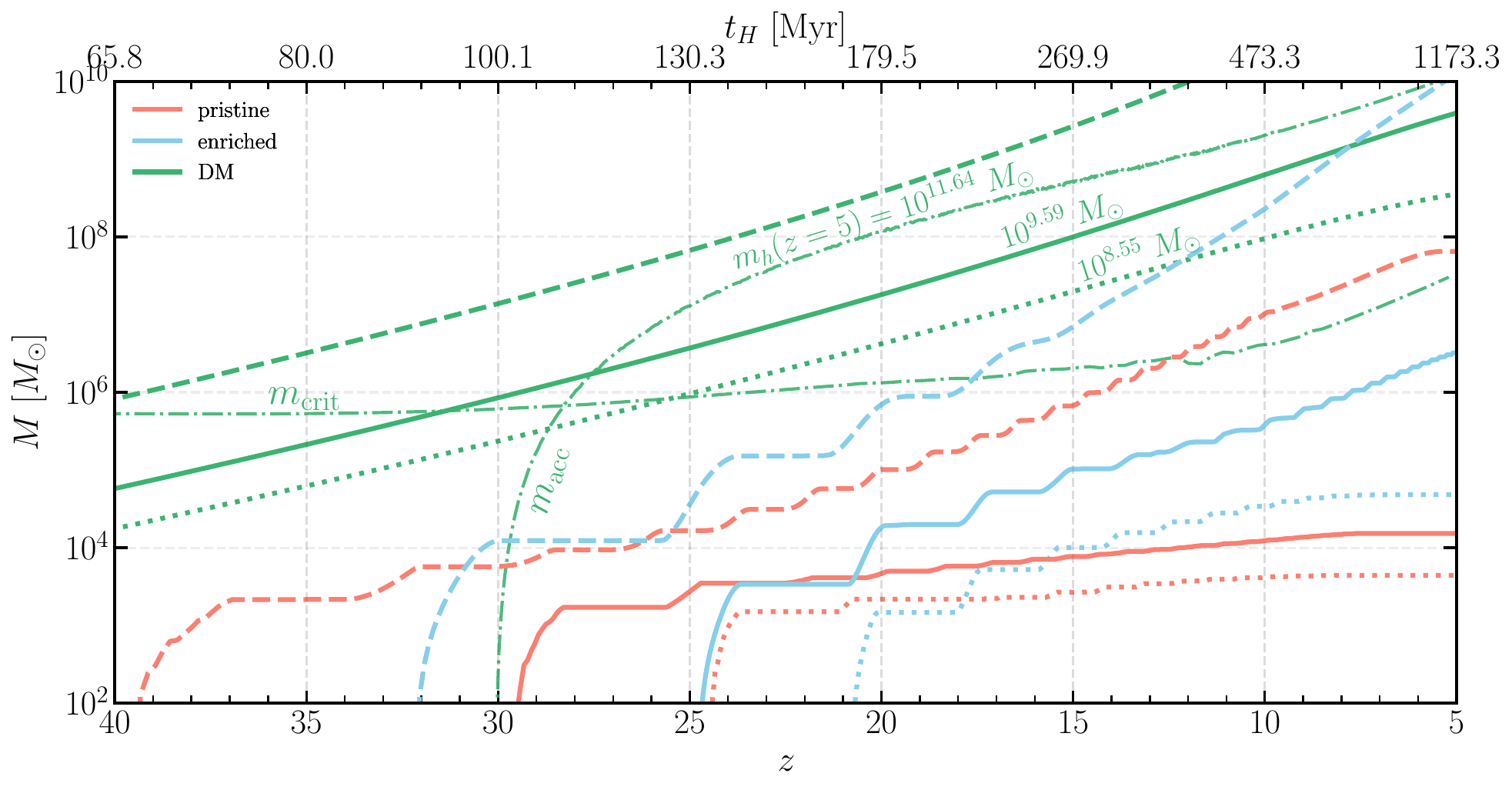}
    \caption{\textbf{Environmental feedback can introduce significant variations in star formation efficiency, depending on the host halo mass.} Evolution of the same $\sim 10^{9.5}\ M_\odot$ halo from Figures~\ref{fig:SFH_ex} and \ref{fig:SFH_reionization_feedback} (solid curves) compared with a more massive $\sim 10^{11.6}M_\odot$ halo (dashed) and a lower mass $10^{8.5}M_\odot$ halo (dotted), with their associated star formation histories (cumulative pristine stellar mass in red and enriched in blue) including all components of the model, both internal and environmental.}
    \label{fig:SFH_Mh_comparison}
\end{figure*}

By suppressing accretion in this manner, where we compute the effects of reionization feedback through the calculation of a global UV background to which a halo is exposed once it begins the star formation process, we are not attempting to model the effects of `self-ionization,' whereby a halo could suppress accretion onto itself by building up a local ionized bubble. In this sense, the assumption underpinning our necessarily simplified approach to modelling reionization feedback is that a local ionized bubble is unable to prevent accretion from flows already nearby the halo (consistent with the picture outlined in \citealp{noh_physical_2014}), and instead the global ionizing background dominates in this feedback regime, wherein ionization and heating of the low-density IGM suppresses the ability for small halos to continue to accrete gas. 

In the upper panel of Figure~\ref{fig:SFH_reionization_feedback}, we show the evolution of the two copies of the same $10^{9.5} M_\odot$ halo from Figure~\ref{fig:SFH_ex} with and without reionization feedback. We also show the associated probability that a halo of this mass evolves in an ionized bubble, the quantity that will ultimately set the relative weightings of these two histories to the total SFRD and other associated quantities. From these histories, it is evident that reionization feedback will significantly affect the evolution of halos smaller than $\sim 10^{10} M_\odot$, and these reionization-affected halos will become increasingly more common after $z\lesssim 15-20$. In the lower panel, we show the associated star formation rates for the halos within and outside ionized bubbles, demonstrating that, on average, the SFRs are suppressed by roughly an order of magnitude as reionization proceeds.

\subsection{A complete galaxy history}\label{sec:external_summary}
Figure~\ref{fig:SFH_reionization_feedback} also includes the enrichment of the IGM, thereby highlighting the additional effects of external feedback processes atop those discussed in Section~\ref{sec:internal_summary} and shown in Figure~\ref{fig:SFH_ex}. By inspection, it is clear that while the early time behavior of this halo is largely unchanged, there are key differences at late times as the IGM enrichment and reionization feedback set in. That is, as the halo's enriched winds begin to penetrate filaments in the IGM (Section~\ref{sec:global_enrichment}), the growth of the pristine reservoir is suppressed, leading to a downturn in the rate of pristine gas inflow to the ISM and associated SFR, even in the absence of reionization feedback. Reionization feedback compounds this effect, and at the latest times, the halo is unable to maintain any pristine star formation, irrespective of the ionization state of its local IGM.

\section{Population effects}\label{sec:pop_quantities}
So far we have focused on the evolution of a single halo (albeit one exposed to large-scale feedback). In what follows, we will discuss the manifestation of these components of our model at a population level, integrating over the star formation histories of a broad range of halos, as we self-consistently couple their individual reservoirs to global thresholds such as the critical halo mass for the onset of star formation and those set by the thermal and chemical states of the IGM.

To do this, we evolve equations (eqs.~\ref{eq:m_ism_ev}-\ref{eq:E_CGM_enr}) from $z=50$ to $5$ with $0.5\ {\rm Myr}$ timesteps in parallel for 1000 halos with masses logarithmically spaced between $10^7M_\odot$ and $10^{14}M_\odot$ at $z=5$. We use the fiducial parameter choices listed in Table~\ref{tab:constants}, namely with a moderately top-heavy pristine IMF (mean mass of $\sim 20 M_\odot$; which characterizes the radiative and SN feedback from the pristine stars) and the `$z$-independent' scaling for the Pop II feedback mass loading from \cite{furlanetto_minimalist_2017} calibrated to high-$z$ luminosity function observations (see Figure~\ref{fig:UVLF}). 

\subsection{Star formation histories}\label{sec:sfh_mult}
In Figure~\ref{fig:SFH_Mh_comparison}, we compare star formation histories computed with our model for halos of three different masses, demonstrating the interplay of different components of the model framework on the star formation histories in both the pristine and enriched reservoirs. By examining the star formation histories of these representative halos, we can build intuition for the typical evolution of a halo in various mass bins and understand the relevant physical processes that determine the relative rates of star formation out to late times.

The most massive of the three halos --- which is two orders of magnitude larger than that shown in Figures~\ref{fig:SFH_ex} and \ref{fig:SFH_reionization_feedback} --- begins its star formation earliest, shortly after crossing the $m_{\rm crit}$ threshold, and it is the only one of the three shown halos that has a high level of ongoing Pop III star formation through the end of our calculations at $z=5$. Indeed, only the most massive halos ($m_h \gtrsim 10^{11} M_\odot$ at $z=5-10$) are able to continue accreting gas as the universe becomes reionized and the IGM heated and enriched. In such halos, which are more resistant to feedback events, mass does not escape and thus local enrichment of the IGM proceeds inefficiently, leading to an elevated late-time level of pristine star formation. In turn, such halos are well above any reionization accretion threshold and thus evolve almost independent of any external processes. 

At lower masses ($m_h\sim 10^9-10^{11} M_\odot$ --- the same halo shown in Figures~\ref{fig:SFH_ex} and \ref{fig:SFH_reionization_feedback}), reionization feedback and IGM enrichment are the dominant mechanisms in determining the long-term star formation history. In these halos, the early star formation behavior is similar to the more massive halo, just delayed by the critical mass criterion. However, at later times, such halos, which sit just below the accretion threshold (eq.~\ref{eq:Mcrit_acc}) have their gas supply limited, resulting in a slowing or quenching of their star formation, in effect `freezing-in' the pristine stellar populations that were born between $z\sim 8-12$. In tandem, these halos are low mass enough that early SN feedback-driven winds enrich their local surroundings and quickly shut off pristine star formation.

\begin{figure*}
    \centering
    \includegraphics[width=0.9\linewidth]{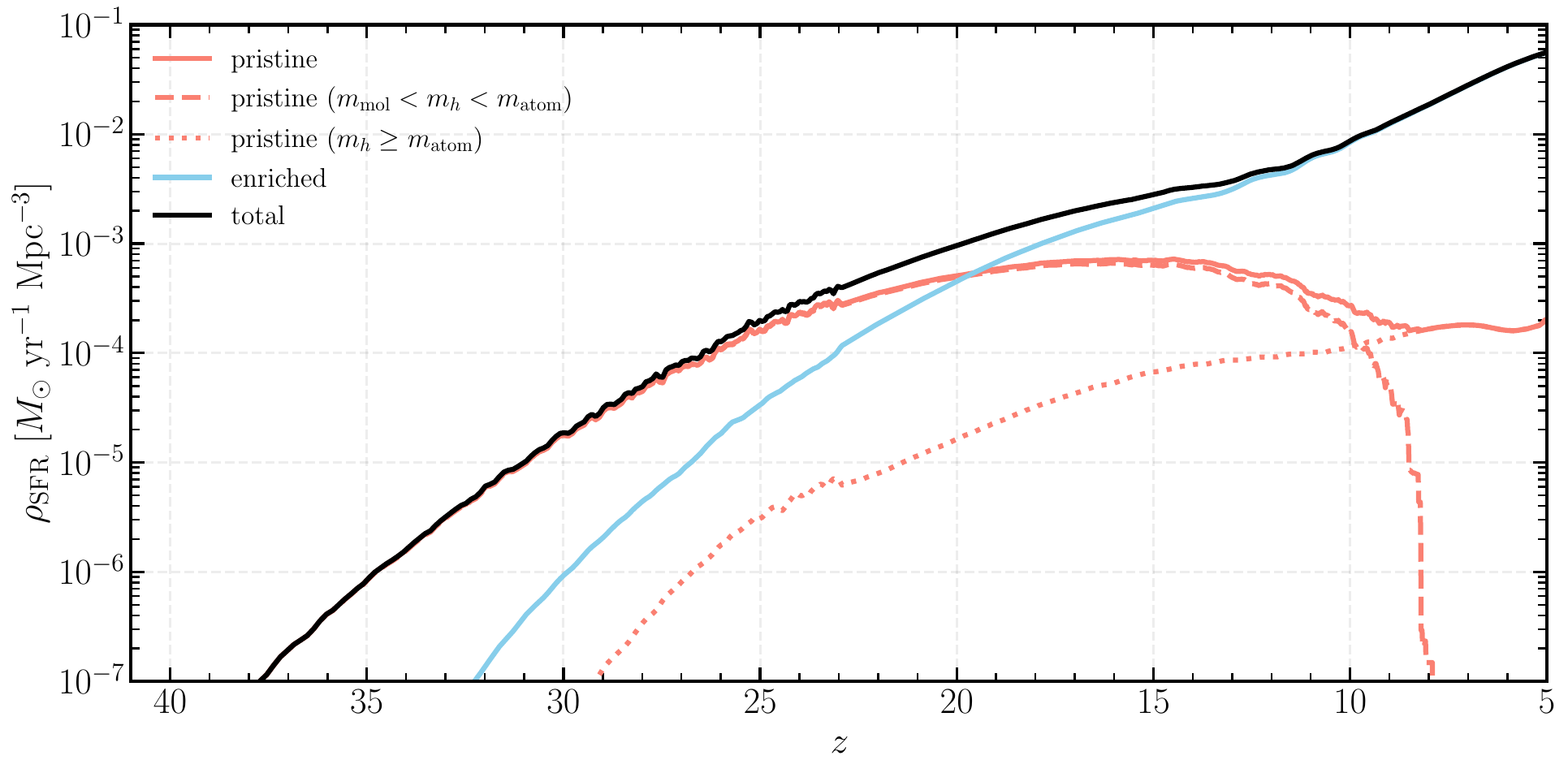}
    \caption{\textbf{A high level of pristine star formation can continue through the EoR, but is limited to the most massive halos (those above the ACT).} The global star formation rate density (SFRD; black) for our fiducial model parameter set, split into the pristine (red) and enriched (blue) components. We also highlight the contribution to pristine star formation from molecular (dashed) and atomic (dotted) cooling halos.}
    \label{fig:SFRD_Mhsplit}
\end{figure*}

\begin{figure*}
    \centering
    \includegraphics[width=0.9\linewidth]{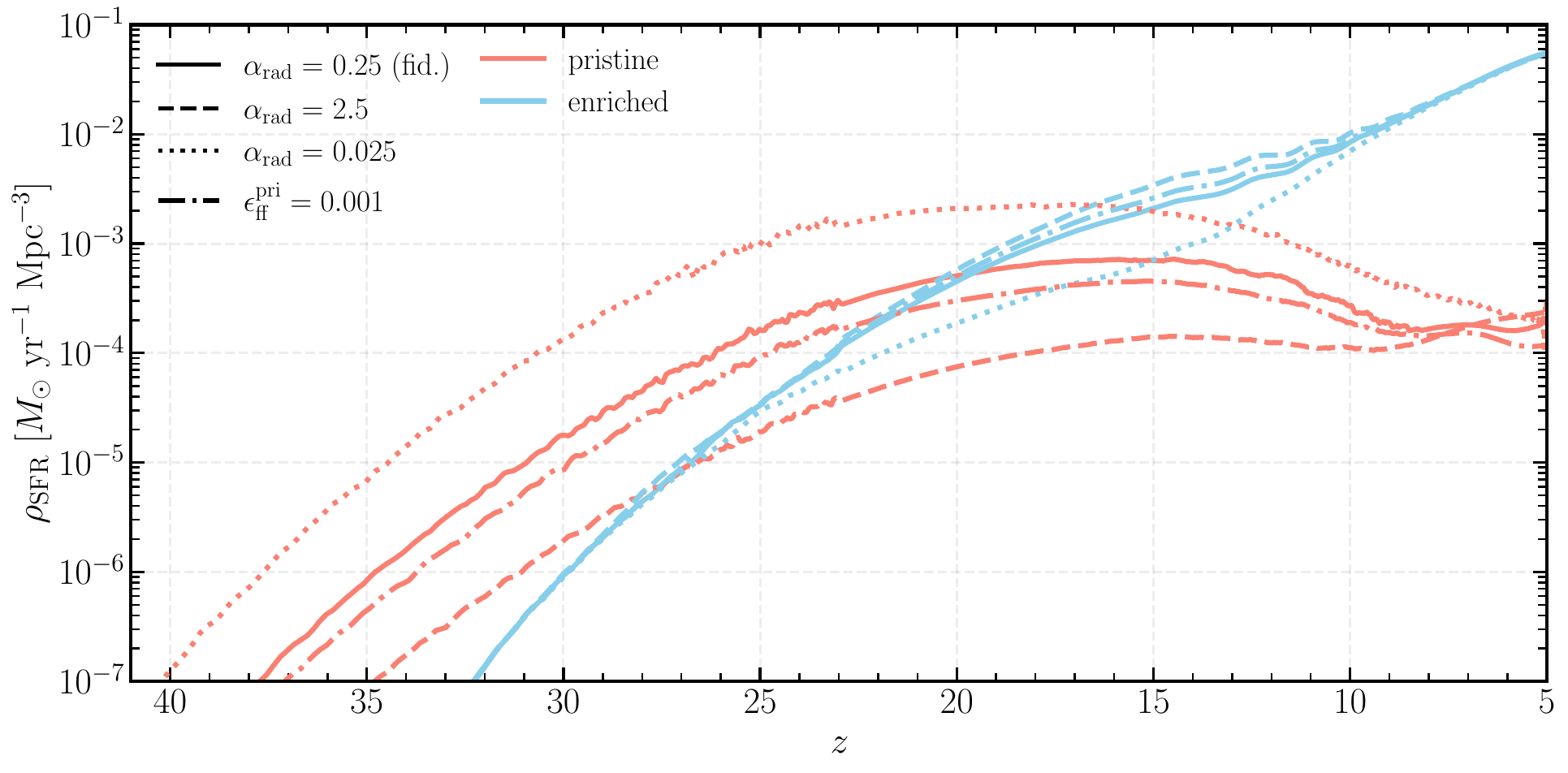}
    \caption{\textbf{The details of star formation on small scales can significantly modulate the level of early-time star formation.} Variation of the SFRD over time with different choices for the fudge factor $\alpha_{\rm rad}$ introduced in eq.~\ref{eq:sfe_max} (linestyles) compared to a case with a fixed $\epsilon_{\rm ff}^{\rm pri} = 0.001$ (dot-dashed).}
    \label{fig:SFRD_alphaRad_variation}
\end{figure*}

At lower masses still ($m_h \lesssim 10^9 M_\odot$), halos are unable to cross the critical star formation threshold until $z\lesssim 25$, and thus form only a single/few burst(s) of stars before reionization feedback quenches their later growth. This results in a relatively higher ratio of pristine to enriched stellar mass at late times, possibly suggesting that such systems, if detectable, could be promising sites of detecting the signatures (or remnants) of pristine star formation.\footnote{While the most massive pristine stars, which are conventionally touted as producing `smoking gun' signatures of Pop III star formation, are likely to have reached the end of their lives by the time of observation ($z\sim 5-10$), the quenching of enriched star formation could result in their chemical imprints on the gas of the ISM being more cleanly preserved, as is explored in \cite{salvadori_first_2012, Kulkarni13, vanni_chemical_2024, sodini_evidence_2024, visbal_chemical_2025}, for example. Estimating the strength of these signatures is, however, beyond the scope of this work.}

From these individual histories, it is clear that the external feedback processes (Section~\ref{sec:external_modifications}) are the dominant determinants of the star formation history of an individual galaxy below $m_h\lesssim 10^{10} M_\odot$, especially at the latest times. In turn, we highlight two qualitative conclusions: (1) reionization feedback sets the \textit{level of total star formation} achieved in a galaxy; and (2) enrichment of the IGM characterizes the \textit{relative levels of star formation in the two phases}, so efficient propagation and mixing of enriched bubbles into a galaxy's local neighborhood would result in a lower level of late-time pristine star formation. Because the global filling fraction of metals in the IGM is low even by $z\sim 5$, the relative levels of star formation in the two reservoirs is dictated by mixing in the CGM (Section~\ref{sec:cgm}) and local enrichment (Section~\ref{sec:global_enrichment}). From Figure~\ref{fig:SFH_reionization_feedback} (specifically comparing the solid and dotted lines), we can see that while mixing begins the process of bifurcating material between the two reservoirs, local enrichment hastens the process (and can even quench pristine star formation in some cases). This process, however, is most efficient in lower mass halos, where outflows are able to more efficiently escape. 

Higher mass halos ($m_h\gtrsim 10^{10} M_\odot$), on the other hand, are relatively immune to reionization feedback and are only mildly affected by local enrichment as they can more efficiently trap outflows.

\begin{figure*}
    \centering
    \includegraphics[width=0.9\linewidth]{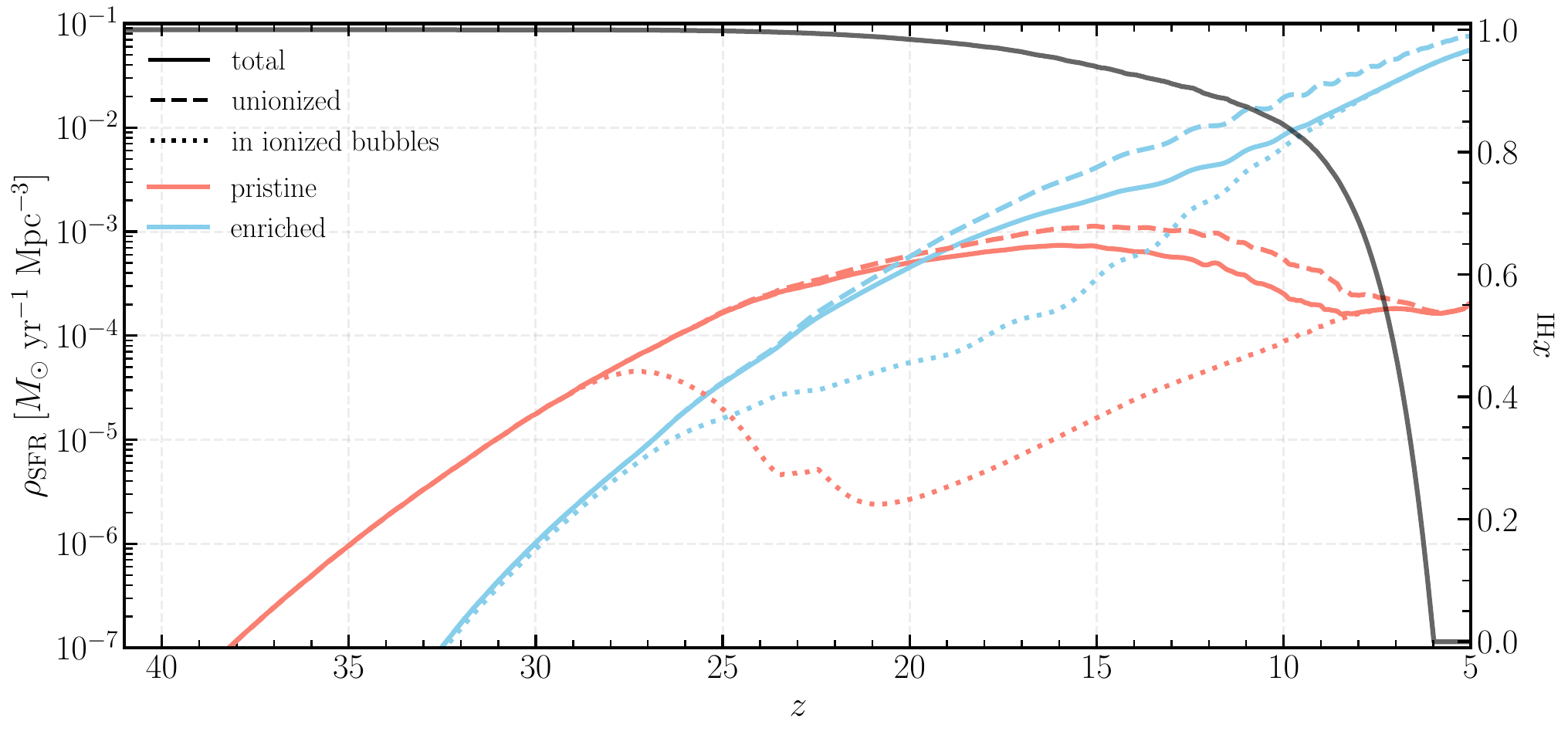}
    \caption{\textbf{Reionization feedback can significantly modulate the global Pop III and II SFRDs.} The global star formation rate density (SFRD; black) for our fiducial model parameter set, split into the pristine (red) and enriched (blue) components, highlighting the contribution from halos in ionized (dotted) and unionized (dashed) regions. The global neutral fraction is shown as a black solid curve, corresponding to the right vertical axis.}
    \label{fig:SFRD_reionization_feedback}
\end{figure*}

\begin{figure*}
    \centering
    \includegraphics[width=0.9\linewidth]{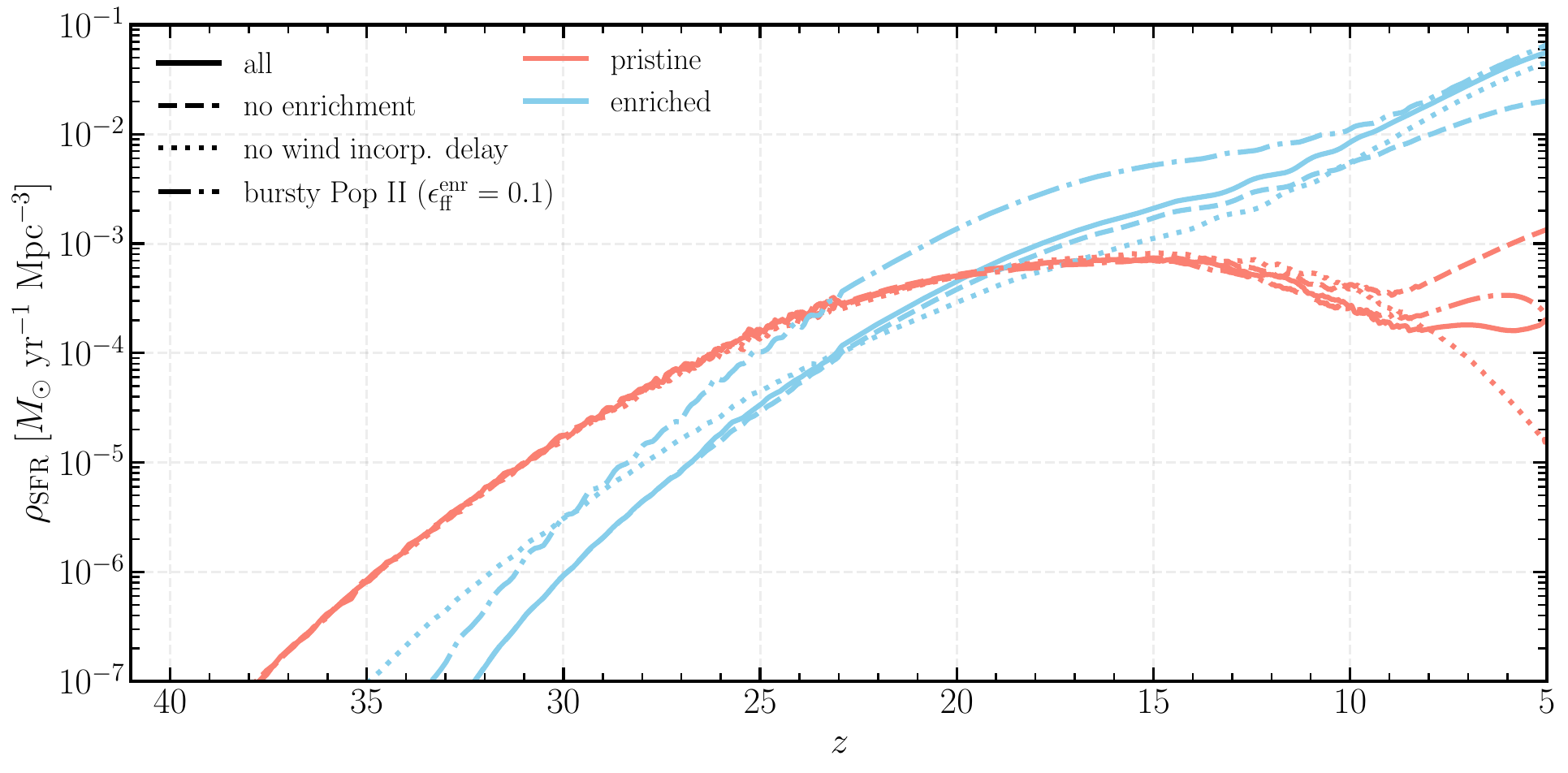}
    \caption{\textbf{The most significant regulator of the late-time pristine star formation rate is the enrichment of the IGM, though the other feedback mechanisms play roles in mostly distinct cosmic epochs.} Variation in the evolution of the pristine (red) and enriched (blue) SFRDs as different feedback components are individually varied (different linestyles), compared to the fiducial model (solid). The model with no IGM enrichment (Section~\ref{sec:global_enrichment}) is shown with a dashed line, a case with no time delay between stellar feedback and the incorporation of winds into the CGM (Section~\ref{sec:PopIII-II_transition}) is shown with a dotted line, and a case where the Pop II SFE is enhanced is shown with a dot-dashed line (Section~\ref{sec:burst_baseline} and Table~\ref{tab:constants}).}
    \label{fig:SFRD_component_comparison}
\end{figure*}

\subsection{The global SFRD}\label{sec:global_sfrd}
In Figure~\ref{fig:SFRD_Mhsplit}, we integrate over the aforementioned halo population to compute the contributions of the two stellar populations to the global SFRD. This summary plot makes evident some of the qualitative conclusions drawn from the representative galaxy histories discussed in Section~\ref{sec:sfh_mult}. That is, the early time SFRD is dominated by pristine star formation and roughly traces the halo growth rate as this era is mostly characterized by when minihalos cross the critical mass threshold in combination with the star formation efficiency (Section~\ref{sec:rad_feedback}). The transition to enriched star formation in the total SFRD does not occur until $z\sim 20$, with this delay a result of the recycling framework discussed in Section~\ref{sec:PopIII-II_transition}. Once the enriched SFRD grows substantially, internal LW feedback suppresses the pristine SFRD in molecular cooling halos, and the pristine SFR in such halos plummets. Then, the shape of the late-time pristine SFRD is initially characterized by the timescales of halos crossing the atomic cooling threshold. In fact, at the latest times (into the EoR, $z\lesssim 8$), \textit{there is no more pristine star formation in molecular cooling `minihalos,' and any ongoing pristine star formation is limited to more massive halos}, where it occurs alongside enriched star formation (similarly to what was seen in the simulations of \citealp{brauer_aeos_2025b, zier_THESAN_2025}). In tandem, the (dominant) global enriched SFRD grows steadily as halos go through cycles of bursty star formation until converging to an equilibrium (albeit still growing) value of $\rho_{\rm SFR}^{\rm enr}(z\sim 5) \sim {\rm few}\times 10^{-2}\ M_\odot\ {\rm yr^{-1}\ Mpc^{-3}}$.

\subsubsection{The Pop III SFRD at early times}\label{sec:early_SFRD}
As was discussed in the preceding section, the SFRD at early times is dominated by molecular cooling minihalos, whose level of star formation is set by radiative feedback in star forming clouds. In Figure~\ref{fig:SFRD_alphaRad_variation}, we highlight this effect by varying our choice of $\alpha_{\rm rad}$, the constant appearing in eq.~\ref{eq:sfe_max}, which ultimately characterizes the integrated star formation efficiency in the pristine phase. From this investigation, it is clear that order of magnitude variations in $\alpha_{\rm rad}$ result in nearly order of magnitude variations in the high-redshift SFRD. We calibrate our fiducial choice by comparing against a model where the star formation efficiency per free-fall time is set to a halo mass-independent value of $\epsilon_{\rm ff}^{\rm pri}  = 0.001$, as is sometimes used in the literature. Despite the significant variations in the level of early star formation with variations in $\alpha_{\rm rad}$, we note that all our models converge to roughly the same late-time value, and thus conclude that any low-redshift predictions are fairly insensitive to this choice. While independent numerical predictions for $\epsilon_{\rm ff}^{\rm pri}$ remain unconverged, this investigation demonstrates that any sensitive probe of early Pop III star formation may yield valuable constraints on this quantity, which can ultimately contribute to our understanding of star formation in pristine clouds.

\begin{figure*}
    \centering
    \includegraphics[width=0.9\linewidth]{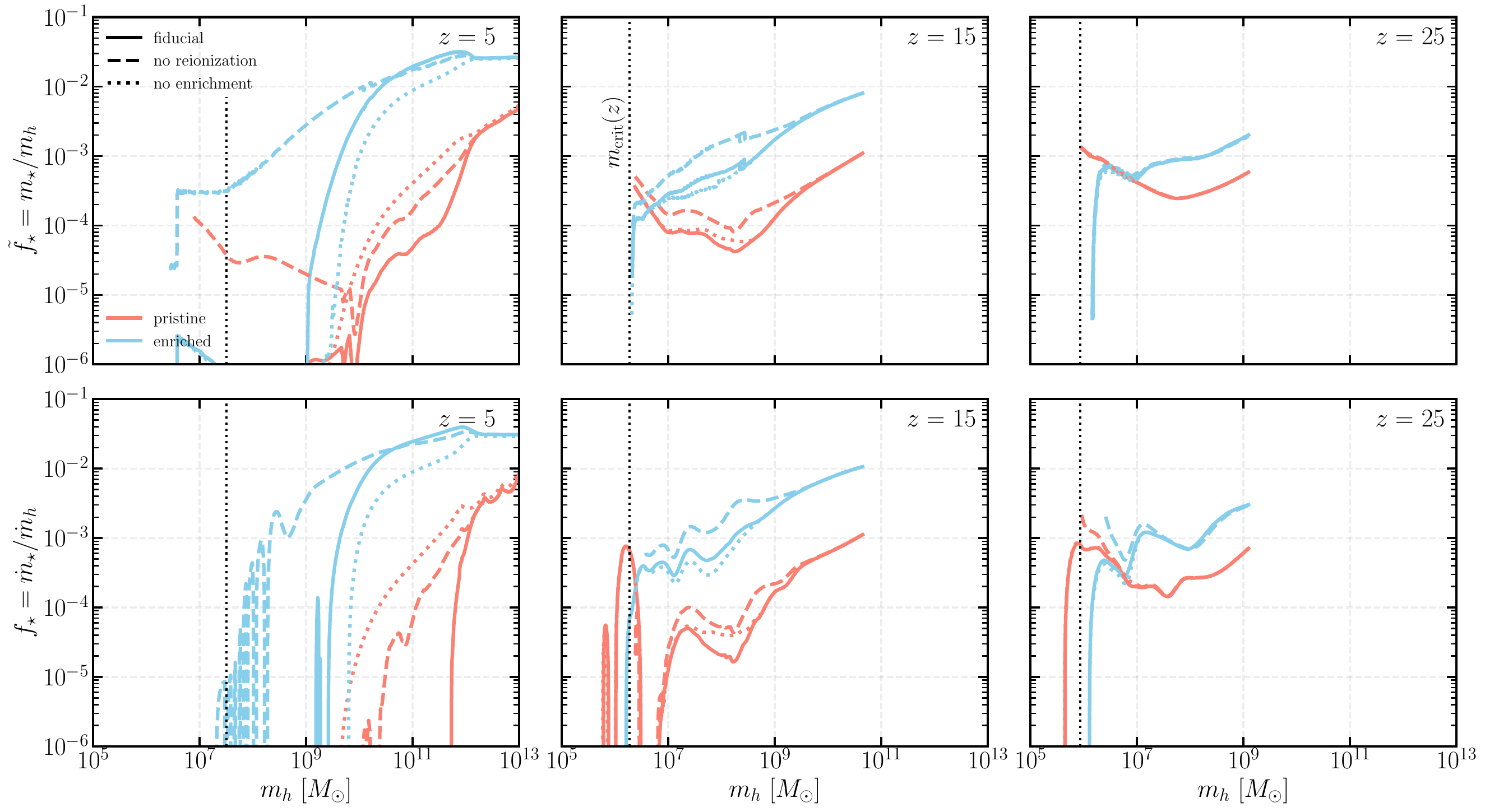}
    \caption{\textbf{At late times, pristine and enriched star formation occur in the same (massive) halos.} \textit{(upper)} The \textit{integrated} pristine (red) and enriched (blue) star formation efficiencies for the fiducial model (solid), a model with no reionization feedback (dashed), and one with no enrichment (maximal pristine star formation; dotted) at $z=5, 15,$ and 25 (three panels, left to right). \textit{(lower) The \textit{instantaneous} star formation efficiencies for the same models as the upper panels.}}
    \label{fig:sfe_evolution}
\end{figure*}

\subsubsection{The Pop III SFRD at later times}\label{sec:late_SFRD}
Here we demonstrate some of the most significant contributions to the late-time Pop III SFRD and highlight areas of uncertainty in parameters underpinning pristine star formation.

First, Figure~\ref{fig:SFRD_reionization_feedback} demonstrates the manifestation of reionization feedback in our model, which begins to affect the SFRD after $z\sim 20$. With the additional effects of clustering modifying the probability that a halo is in an ionized bubble (as discussed in Section~\ref{ssec:reion_feedback}), the relative contributions to the global SFRD track the neutral fraction, as the SFRD moves from tracing that of halos in neutral regions to those in ionized regions at $z\lesssim 10$. Because star formation at the latest times is dominated by massive, atomic cooling halos which are relatively unaffected by the accretion suppression induced by IGM heating, at the latest times the two SFRD histories (ionized and neutral) converge to the same value. However, despite the substantial effect of reionization on individual small halos, it does not meaningfully affect the integrated SFRD because the photoheating feedback does not become important until $z \lesssim 10$, once the global neutral fraction has dropped considerably. We also note that the qualitative conclusions around reionization feedback are relatively insensitive to our choice of observationally-uncertain parameters, such as the production efficiency of ionizing photons $\xi_{\rm ion}^{\rm (i)}$ or the associated escape fraction $f_{\rm esc}^{\rm (i)}$. That is, if reionization were to proceed more rapidly (for example ending by $z\sim 9$ as suggested by \citealp{munoz_reionization_2024}), then the global SFRD would deviate from the unionized track (dashed lines) and converge to the ionized case (dotted) correspondingly earlier, but the ultimate value would be unchanged.

To understand the processes shaping the late-pristine SFRD in more detail, we  use Figure~\ref{fig:SFRD_component_comparison}, where we have independently varied or removed the dominant components of the model. First, the recycling time delay between the injection of stellar winds and their incorporation into the CGM introduces a delay in the onset of enriched, Pop II star formation, though the curves largely converge to similar values at late times. However, because of this elevated level of early enriched star formation, IGM enrichment proceeds more efficiently and the late-time pristine SFRD is more efficiently suppressed. Relatedly, a model run with no IGM enrichment (either by local wind bubbles or from global enrichment) demonstrates a \textit{maximal} late-time estimate of pristine star formation elevated by a factor of $\sim 3-4$ over the fiducial value. Finally, the dot-dashed curve highlights the sensitivity (or lack thereof) of the Pop III SFRD to the efficiency with which enriched gas is converted into stars. As is discussed in Section~\ref{sec:observables}, the baseline Pop II model is unable to adequately reproduce observed luminosity functions at the earliest times, and one such modification to accommodate this underprediction is to enhance the efficiency (and thus burstiness) of early star formation. From inspection of Figure~\ref{fig:SFRD_component_comparison}, we find that despite a nearly order of magnitude boost to the enriched SFRD induced by an enhanced SFE, there is minimal effect on the associated pristine SFRD, so we conclude that our results are relatively insensitive to this effect.

Taken together, these curves provide a range of estimates for the late-time pristine SFRD, yielding an optimistic level of $\mathcal{O}(10^{-3})\ M_\odot\ {\rm yr^{-1}\ Mpc^{-3}}$ in the limit that enrichment and mixing proceed inefficiently in the IGM and/or if the Pop III SFE is higher than we have fiducially assumed. Compared with the enriched SFRD, at the latest times, we expect Pop III star formation to constitute $< 10\%$ of the total star formation.

\subsection{Star formation efficiencies}\label{sec:SFE}
The upper row of Figure~\ref{fig:sfe_evolution} shows the \textit{integrated} star formation efficiency (SFE$=m_\star^{\rm (i)}/m_h$) over time computed from the fiducial model. From this we see many of the same trends as have been discussed in the preceding sections. Namely, at early times, pristine star formation is most prominent in halos close to the molecular cooling threshold, while enriched star formation occurs only in more massive halos that have already undergone earlier star formation episodes. Between $z=25$ and $15$, however, enriched star formation takes over as the dominant star formation mode in most halos, a trend which continues to the latest times. In fact, we see a dip in the pristine SFE at around $m_h\sim 10^9 M_\odot$, reflecting the effect of  local enrichment of the IGM on efficiently enriching inflows in such halos. At this point, reionization feedback has already begun to suppress star formation in the lowest mass halos, and stellar mass buildup does not proceed in halos below $\sim {\rm few}\times 10^6M_\odot$. By $z\lesssim 10$, star formation is quenched in halos below $\sim 10^9 M_\odot$ and the efficiency of local enrichment of the IGM determines the level of pristine star formation that can persist. Indeed, as is seen in the simulations of \cite{mead_aeos_2025}, only the most massive halos are able to retain their metals. When coupled with inefficient mixing in the CGM, this limits pristine star formation to only occurring in such systems \citep{venditti_needle_2023}. These limits suggest that Pop III integrated SFEs on the order of ${\rm a\ few}\times 10^{-3}$ and stellar mass ratios of roughly 10\% are typical in the most massive halos. The integrated SFE provides a good proxy for identifying systems wherein tracers of Pop III remnants, such as unique chemical abundance patterns associated with their SNe, might be found. For example, our results demonstrate that low mass halos ($m_h\lesssim 10^{9} M_\odot$) that persist in neutral regions to the end of reionization could host Pop III relics with stellar mass ratios of $1-10\%$, reinforcing the idea that such systems, some of which may appear as damped Ly$\alpha$ absorbers are promising sites for detecting the imprints of Pop III star formation.

The lower panels in Figure~\ref{fig:sfe_evolution} show the \textit{instantaneous} SFE ($f_\star=\dot{m}_\star^{\rm (i)}/\dot{m}_h$) predicted by the model, highlighting the sites of ongoing star formation. In this case, we see that independent of the local reionization history, pristine star formation only persists in massive ($m_h \gtrsim 10^{10} M_\odot$) systems. The range of halos in which we expect to find such stars is instead sensitive to the local enrichment history for a particular halo, but the highest level of pristine star formation is seen in the most massive halos.
If such stars forming out of pristine gas in atomic cooling halos are born with a top-heavy IMF, then in principle we would expect to see signatures of these stars in deep spectra of these luminous galaxies, though exploring the strength of these signatures is beyond the scope of this work.

\section{Comparison to past work}\label{sec:past_work}
When and where Pop III star formation occurs is a question that has been considered in a number of other studies. In this section, we briefly highlight the similarities and differences between this model and others in the literature.

The semi-analytic models that attempt to follow Pop III star formation down to the end of the reionization era include those introduced in \cite{visbal_self-consistent_2020}, \cite{liu_when_2020}, \cite{hartwig_public_2022}, and \cite{ventura_semi-analytic_2024}. While the detailed parameterization of star formation, feedback, and reionization differ between these works, the fundamental architecture of all these models is the same: they predict Pop II and III SFRDs by analytically and numerically painting star formation onto DM halos modeled with N-body cosmological simulations. By post-processing star formation histories in a simulated box, for which spatial information and halo merger histories are available, these authors can account for inhomogeneous effects such as LW feedback, IGM enrichment, and reionization, which cannot be followed in as much detail in a comparatively more analytic model like the one introduced here. Despite the detailed differences between these models, they all roughly converge to a late time Pop III SFRD of $\sim 10^{-5}-10^{-4}\ M_\odot\ {\rm yr^{-1}\ Mpc^{-3}}$, which is a factor of a few to an order of magnitude smaller than the value we find in our fiducial model here. While choices in the exact parameterization of the critical mass or star formation efficiency will introduce order unity variations in the SFRD, this difference is largely due to the assumptions in our underlying star formation model. That is, by allowing for Pop III star formation to persist beyond the first star formation episode in a halo --- as a result of inefficient mixing within the halo and inefficient incorporation of feedback-driven winds into the IGM --- we find a higher Pop III SFRD even at the end stages of reionization. 

Several groups have leveraged cosmological simulations to tackle this question as well, most recently with an explicit focus on the effects of inhomogeneous mixing and radiative feedback on late-time Pop III star formation (see e.g., \citealp{pallottini_simulating_2014, sarmento_following_2017, sarmento_following_2018, jaacks_legacy_2019, sarmento_following_2019, venditti_needle_2023, sarmento_importance_2025}). Because these simulations are often resolution-limited at the earliest times, when the Pop III SFRD will be dominated by star formation in the more abundant but smaller minihalos, it is most instructive to focus any comparison on the late-time SFRDs they predict. On average, these simulations show Pop III SFRDs $\sim 10^{-4}-10^{-3}\ M_\odot\ {\rm yr^{-1}\ Mpc^{-3}}$ at $z\sim 5-7$, more in line with the results seen in this work, and consistent with our findings that inefficient mixing can leave a substantial pristine gas reservoir available to form Pop III stars out to the latest times, but that such gas is limited to \textit{only the most massive halos}. 

These comparisons highlight a few key characteristics of our model. First, ours is the first semi-analytic model that attempts to characterize the internal metal mixing process in any detail and allows for continued Pop III star formation \textit{even after the first star formation episodes are complete}. In turn, the analytic nature of our model allows us to efficiently and robustly track star formation in halos across a range of mass scales from early times through the end of reionization, providing a more comprehensive characterization of the Pop III SFRD and integrated star formation efficiency across cosmic time. 

\section{Observational signatures}\label{sec:observables}
In this section, we will summarize some basic observational quantities that this model enables us to predict regarding the potential detectability of pristine, Pop III systems over time. 

\subsection{Luminosity functions and luminosity density}
In order to determine the appropriate form of the feedback mass-loading factor $\eta$ for Pop II stars, we calibrate our model to observed UV luminosity functions. Given the halo masses and associated UV luminosities (mapped from the star formation rates as $\mathcal{L}_{\rm UV} = \dot{m}_\star^{\rm enr}/\kappa_{\rm UV}$), the luminosity function is given by an integral over the halo mass function weighted by the luminosity distribution:
\begin{equation}
    \label{eq:uvlf}
    \Phi(M_{\rm UV}) = \int n(m_h)P(\log_{10}\mathcal{L}(M_{\rm UV})|m_h)dm_h,
\end{equation}
where $n(m_h)$ is the halo mass function and $P(\log_{10}\mathcal{L}|m_h)$ is the as-yet unspecified luminosity distribution around a mean halo mass. From inspection of the luminosity distributions produced by the model, we find that $P(\log_{10}\mathcal{L}|m_h)$ is best described by the asymmetric exponentiated Weibull distribution (rather than the lognormal distribution that is conventionally assumed; see Appendix~\ref{app:lum_dist} for more details):
\begin{equation}
    \label{eq:weibull}
    P(\log_{10}\mathcal{L}|m_h) = \frac{ac}{d}\big[1-\exp(-x^c)\big]^{a-1}\exp(-x^c)x^{c-1},
\end{equation}
where $x = (\log_{10}\mathcal{L} - b)/d$ and $a,b,c,d$ are shape and location parameters that depend on the halo mass $m_h$. To infer the luminosity function from our model at a particular redshift, we then first fit for the parameters $a,b,c,d$ as a function of halo mass and then integrate over the halo mass distribution (eq.~\ref{eq:uvlf}). Here we find that deviations from lognormal occur at low halo masses and high redshift, so the procedure we carry out has the most significant effect at the faint end at $z\gtrsim 8$.

\begin{figure}
    \centering
    \includegraphics[width=0.9\linewidth]{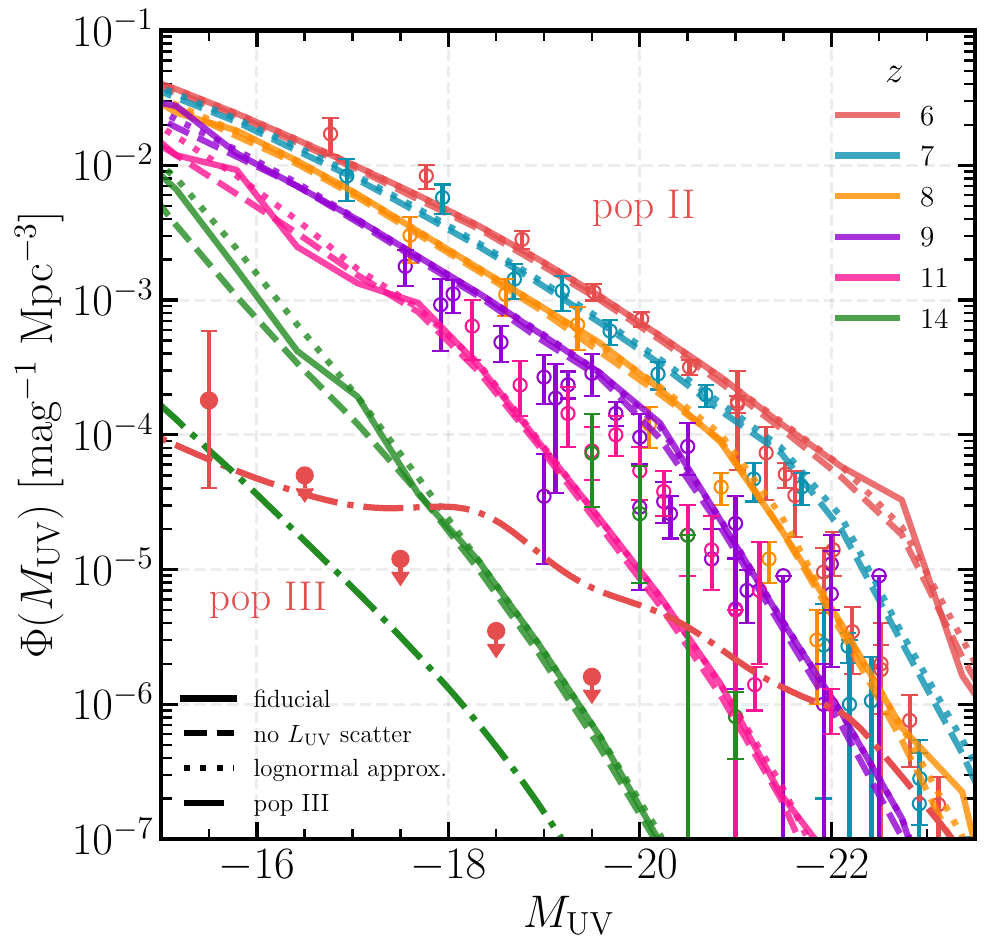}
    \caption{\textbf{The Pop II model agrees well with observed LFs from $z\sim 6-10$, but underpredicts the data at higher $z$.} Luminosity function predictions from the fiducial model for Pop II stars for $z\sim 6-14$ (different colors) compared with a compilation of observational data (open colored points; \citealp{bouwens_new_2021, harikane_goldrush_2022, varadaraj_bright_2023,  casey_cosmos-web_2024, finkelstein_complete_2024, donnan_jwst_2024}). Different linestyles correspond to a case with no scatter in the $L_{\rm UV}-m_h$ relation (dashed) and a case where we account for the scatter with a lognormal distribution (dotted; see Appendix~\ref{app:lum_dist} for more context). We also include an estimate of the Pop III LF at $z=6$ (14) in red (green) with a dot-dashed linestyle along with the tentative detection and upper limits presented in \cite{fujimoto_glimpse_2025} (filled circles). Note that the overprediction at the bright end is expected as we do not include the effects of dust attenuation in these LF estimates.}
    \label{fig:UVLF}
\end{figure}

\begin{figure}
    \centering
    \includegraphics[width=0.9\linewidth]{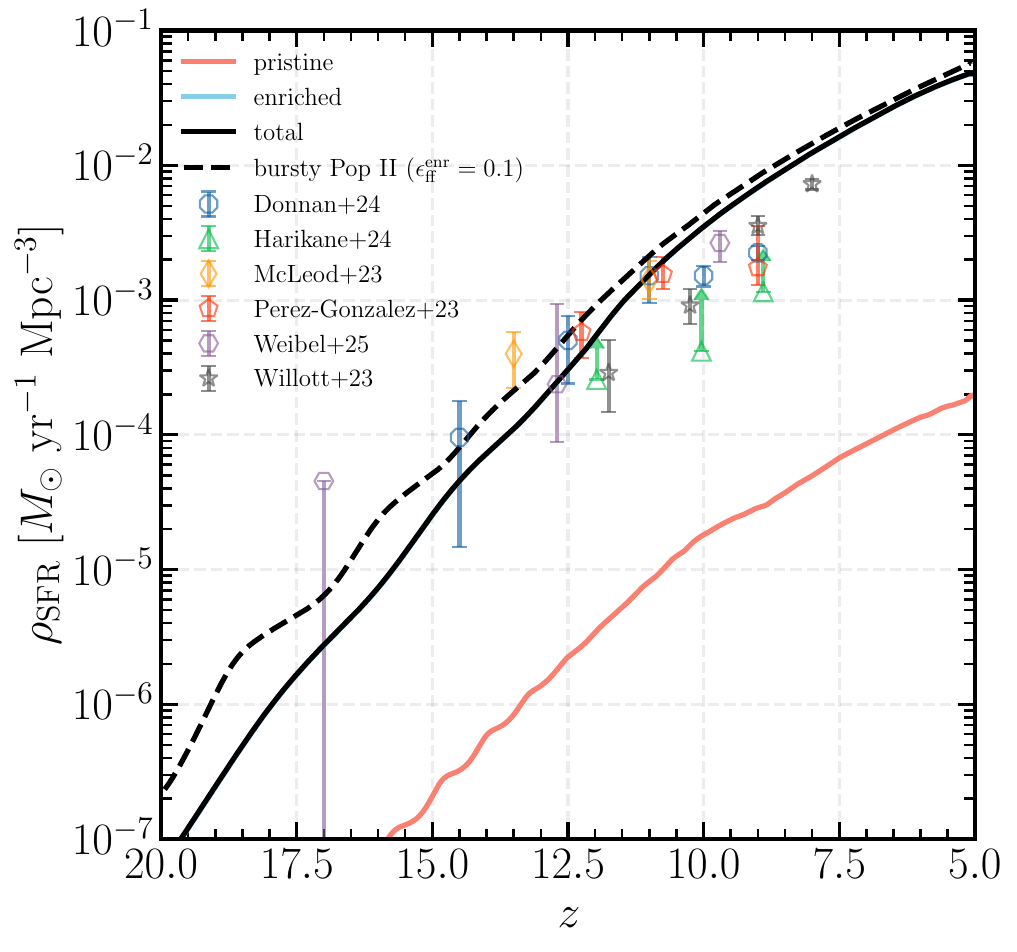}
    \caption{\textbf{The Pop II model provides excellent agreement with the observed star formation rate density over time for $z\lesssim 12.5$.} Magnitude-limited SFRD predictions from the fiducial model for Pop II stars averaged over 30 Myr for $z\sim 5-20$ (different colors) compared with a compilation of JWST-based observational inferences converted from the UV luminosity density with the same numerical factor, $\kappa_{\rm UV}^{\rm enr}$ (colored points; \citealp{mcleod_galaxy_2024, perez-gonzalez_life_2023, willott_steep_2024, harikane_jwst_2025, donnan_jwst_2024, weibel_exploring_2025}). We also show a case where we have enhanced the Pop II SFE by a factor of 10 to $\epsilon_{\rm ff}^{\rm enr} = 0.1$, demonstrating one mechanism to enable better agreement at early times.}
    \label{fig:sfrd_obs}
\end{figure}

From this estimate, we demonstrate that the base Pop II model provides excellent agreement with UVLFs from $z\sim 6-10$ by construction, but underpredicts the abundance of systems at higher $z$, a problem that has been studied extensively in the literature (see e.g., \citealp{mason_brightest_2023, mirocha_balancing_2023, inayoshi_lower_2022, dekel_efficient_2023, ferrara_stunning_2022, hegde_hidden_2024, donnan_no_2025, somerville_density_2025}).\footnote{The disagreement at the bright end at lower redshift is simply because we have not included the effects of dust attenuation in these luminosity function estimates, a component that we defer to later work.} This discrepancy is unsurprising, as we have made no attempt to modify the models (aside from including a physically-motivated model for bursty star formation) from the pre-JWST case. In particular, we find that a redshift-independent feedback mass loading parameter $\eta^{\rm enr}$ is necessary to reproduce the evolution of the observed luminosity functions (as was previously seen in \citealp{furlanetto_minimalist_2017, mirocha_prospects_2020, donnan_no_2025}, for example). However, we do note here that this brief investigation demonstrates that SN time delay-driven bursts do not provide sufficient variability in the $m_h-L_{\rm UV}$ alone, at least for the chosen star formation efficiency of $\epsilon_{\rm ff}^{\rm enr}=0.015$. As is argued in \cite{somerville_density_2025} (and the references therein; e.g., \citealp{kim_modeling_2018, lancaster_star_2021, menon_interplay_2024}), at the high densities we expect for galaxies at early times (see \citealp{williams_star_2025} for a recent simulation highlighting such conditions), the star formation efficiency may rise dramatically, which would in turn increase the strength of burstiness in higher-redshift halos and thus boost our estimates of the UVLF (consistent with recent JWST measurements of bursty star formation at early times; e.g., \citealp{clarke_star_2024, carvajal-bohorquez_stochastic_2025}). In the limit that the density (and thus the SFE per free fall time) is increased, a galaxy will `overshoot' its equilibrium star formation level by an amount modulated by $\epsilon_{\rm ff}$. In turn, the feedback strength will be increased and thus cycles of bursty star formation will proceed for longer (at a fixed halo mass) before converging to the equilibrium solution (see \citealp{furlanetto_bursty_2022} for a detailed discussion of this effect). However, such a time- or density-dependent calibration of our Pop II star formation efficiency to the UVLF is beyond the scope of this work and is worth an independent investigation of its own. Instead here we simply emphasize that while such modifications may be likely, they will \textit{not} affect any of the conclusions about the late-time levels of Pop III star formation, as was seen in Figure~\ref{fig:SFRD_component_comparison}. 

In kind, we compare the predicted pristine, Pop III-only contribution to the UVLF at $z=6$ from our model with the tentative detection from \cite{fujimoto_glimpse_2025} and find good agreement, \textit{though we have not calibrated our Pop III models to any of these observations}. We note that the discrepancy between our model estimate and the GLIMPSE upper limits at the bright end is reasonable and can be attributed to a number of sources. For example, by $z\sim 6$, all pristine star formation occurs in massive halos, which host dominant metal-enriched stellar populations. Subject to assumptions regarding the dust geometry, even the Pop III component could be sensitive to dust attenuation and in turn the associated UV luminosities could be suppressed. In addition, on the observational front, surveys like GLIMPSE targeted distinct regions or clumps that showed unambiguous photometric signatures of star formation from a top-heavy IMF in a low-metallicity environment. In the event that such clumps form in the outskirts of larger metal-enriched systems, as we assume here, then in practice it may be difficult to observationally disentangle the Pop II and III components of a more massive host system and some of the sites of pristine star formation that we identify in our theoretical forecast may thus be missed.

In addition, it is likely that the stellar IMF (and thus the mass-to-light ratio of the stellar populations) displays some metallicity dependence. As a result, we expect that the Pop II IMF will gradually transition from the more top-heavy IMF applied to pristine gas to the conventional Salpeter IMF at later times, though some studies, such as \cite{cueto_astraeus_2025}, note that any decreases to the mass-to-light ratio are effectively neutralized by a decreased star formation efficiency, leaving the overall UVLF essentially unchanged. Others, such as \cite{menon_interplay_2024}, suggest that efficient star formation could coexist with a more top-heavy IMF in low-metallicity star clusters. Given this theoretical uncertainty and degeneracies inherent to such an exploration, especially in conjunction with the uncertainty in the Pop II SFE, we defer these modifications to future work.

In Figure~\ref{fig:sfrd_obs}, we compare our magnitude-limited fiducial model to observed estimates of the SFRD made with JWST, finding excellent agreement at $z\lesssim 11$. We also show a case where we have uniformly (over halo mass and redshift) increased the star formation efficiency in enriched gas by a factor of 10 to $\epsilon_{\rm ff}^{\rm enr} = 0.1$, which introduces a higher level of burstiness and provides a better fit to the SFRD out past $z\sim 14$.\footnote{Note that the oscillations seen in these curves are the result of an increased level of burstiness in individual halos' SFHs and are mostly an artifact of the modelling approach. That is, in the magnitude-limited SFRD, where fewer halos are contributing to the integral, fluctuations in the SFR (which appear more periodic than stochastic in our model) can manifest as oscillations in the global SFRD, so when burstiness is heightened, these features are more pronounced.} 

\begin{figure}
    \centering
    \includegraphics[width=0.9\linewidth]{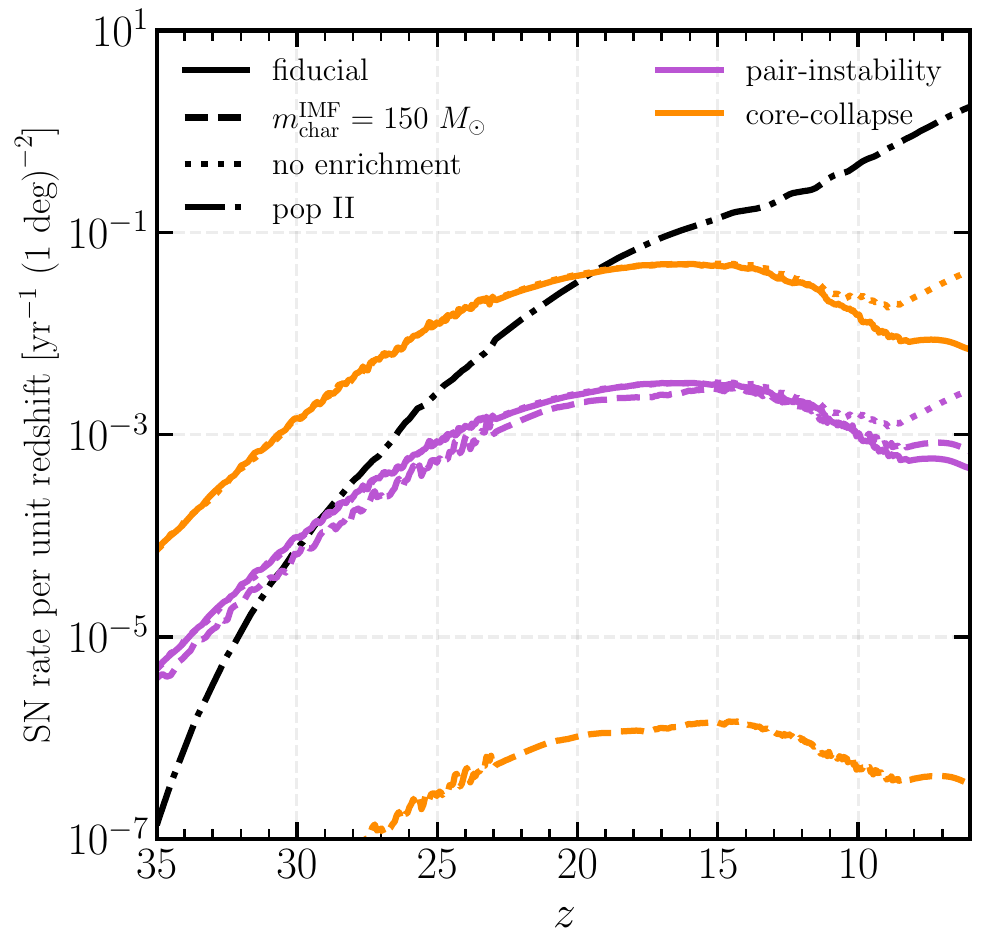}
    \caption{\textbf{Despite there being more PISN progenitors with a very top-heavy IMF, feedback-driven suppression of the global SFRD results in a \textit{lower} late-time PISN rate compared to a more modest, but still top-heavy, IMF.} Cumulative Pop III SN rates for PISNe (purple) and CCSNe (orange) in three different model variations: the fiducial case (solid), a very top-heavy IMF (dashed), and the case with a no IGM enrichment (maximal Pop III star formation; dotted). For comparison, we also show the Pop II CCSNe rate as a black dot-dashed curve.}
    \label{fig:SN_rate_deg}
\end{figure}

\subsection{Transients}
If the stars that form out of pristine gas have a more top-heavy IMF than we infer locally, it is likely that some of these stars will be born in the $140-260M_\odot$ range and could thus end their lives in superluminous pair-instability SNe (PISNe). Such SNe are thought to be nearly two orders of magnitude more luminous than the comparatively more common core-collapse end-of-life behavior (e.g., \citealp{Fryer01, heger_how_2003, kasen_pair_2011}). In the era of deep (and soon wide) IR surveys, the rates of PISNe that we detect (or do not) can provide a sensitive probe of the high-redshift IMF and SFRD. To this end, predictions for the rates we might expect --- a natural extension of the model described above --- are especially useful to help guide the design of our current and next-generation surveys.

With the SFRD computed with our model framework, we can straightforwardly estimate the transient rate as
\begin{equation}\label{eq:transient_rate}
    \frac{d^2N}{dt_{\rm obs}d\Omega_{\rm obs}}(z) = \frac{\eta_{\rm IMF}}{1+z}\frac{d^2V}{dzd\Omega_{\rm obs}}\rho_{\rm SFR}(z),
\end{equation}
where $d^2V/dzd\Omega$ is the differential comoving volume element (determined by our cosmology) and the factor of $1/(1+z)$ accounts for cosmological time dilation. We compute the factor describing the mean number of progenitors per unit stellar mass by integrating over the IMF
\begin{equation}\label{eq:eta_IMF}
    \eta_{\rm IMF} \equiv \frac{\int_X m\phi(m) dm}{\int m \phi(m) dm},
\end{equation}
where $X$ is the progenitor mass range for the transient of interest (e.g., $140-260M_\odot$ for PISNe and $8-40 M_\odot$ for CCSNe). This framework makes evident that the rates of massive stellar SNe that we might expect are determined not only by the underlying IMF, but also by the levels of star formation achieved in the model.

Carrying out this calculation with our fiducial model parameters yields the SN rates shown in Figure~\ref{fig:SN_rate_deg}. From this, we note an important observation --- even though a more top-heavy IMF (such as that shown in the dashed line in the Figure) produces more PISN progenitors per unit star formation, the feedback effects of this more massive stellar population (both in setting the feedback timescales and in characterizing the IGM enrichment; Sections~\ref{sec:PopIII-II_transition} and \ref{sec:global_enrichment}, respectively), results in a marginally lower overall PISN rate (though the ratio of PI/CCSNe in this case is much higher). Indeed, because the radiative and SN feedback effects are so strong in this limit, the star formation efficiency is lower and IGM enrichment proceeds more rapidly, leading to an overall lower level of pristine star formation. 

\begin{figure}
    \centering
    \includegraphics[width=0.9\linewidth]{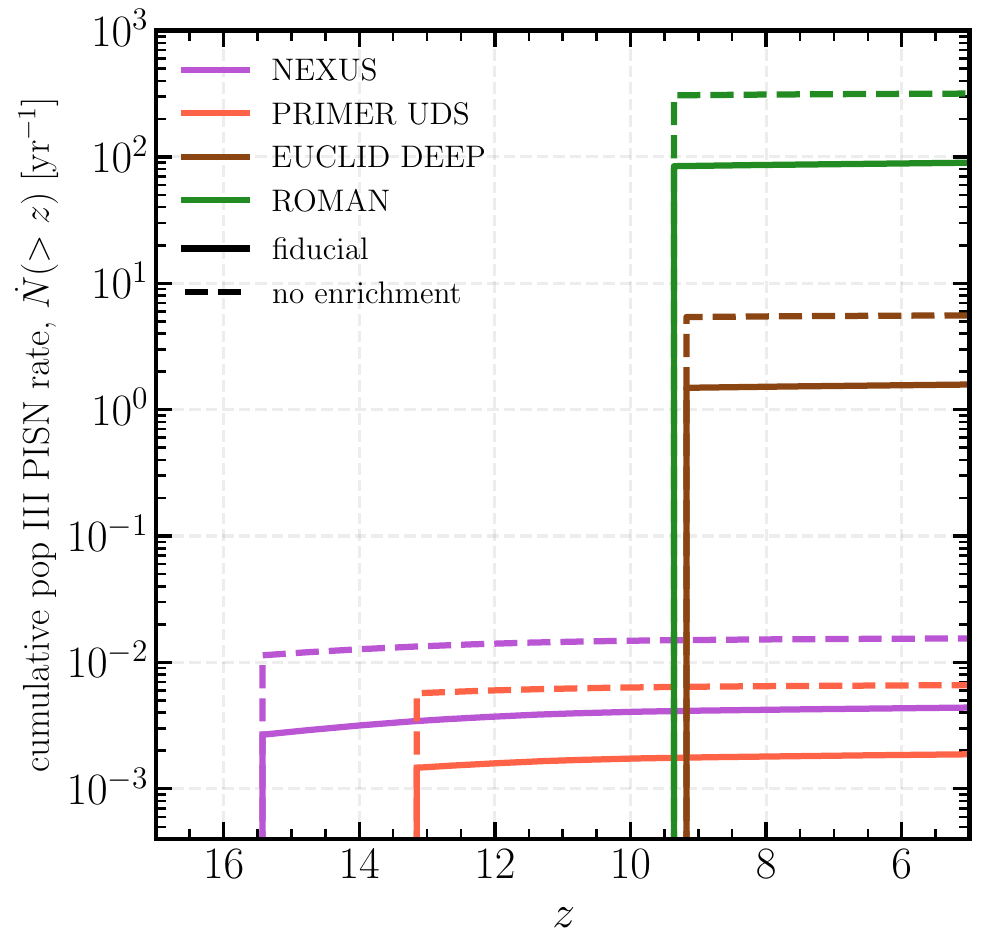}
    \caption{\textbf{Next-generation wide-field instruments will be necessary to have any hope of detecting Pop III PISNe.} The cumulative Pop III PISN rate for four different observing programs with the fiducial and maximal Pop III models (line styles): NEXUS (purple; \citealp{shen_nexus_2024}), PRIMER (red; \citealp{dunlop_primer_2021}), Euclid deep (brown; \citealp{euclid_2022}), and the Roman High Latitude Survey (green; \citealp{wang_HLS_2022}).}
    \label{fig:SN_rate_survey_comp}
\end{figure}

From this Figure, we also note that, even in the limit where the pristine star formation rate is large (i.e., where IGM enrichment is inefficient and pristine star formation proceeds at a higher level to late times), the effect on the PISN rate is modest --- at best, we expect $0.01$ PISNe per year per square degree (per unit redshift). To make this observation more concrete, in Figure~\ref{fig:SN_rate_survey_comp}, we integrate the preceding curves over redshift and show the cumulative Pop III PISN rate for four different current and forthcoming observing programs, folding in their respective survey areas and magnitude limits. In what follows, we use the PISN light curves provided in \cite{kasen_pair_2011} and choose the most luminous stellar model (their `R250'), such that we are demonstrating \textit{an optimistic estimate} of the PISN rate that we expect to detect in any of these surveys. Despite choosing the most optimistic light curve, this Figure demonstrates that it is unlikely to detect Pop III PISNe with any JWST observing program, even in a wide area survey such as NEXUS \citep{shen_nexus_2024}. However, with next generation wide-area space telescopes and programs such as the Roman High Latitude Survey (2277 ${\rm deg}^2$, magnitude limit of 26.5) and the Euclid deep survey (40 ${\rm deg}^2$, magnitude limit of 26.4), the rates rise considerably and a detection becomes more likely (though is still subject to a number of uncertainties, such as the Pop III IMF, SN light curve modelling, etc., and would be overshadowed by the CC SN rate, unless Pop III stars form with an extremely top-heavy IMF). We note that even a considerable boost in the level of star formation (from inefficient enrichment) does not significantly increase the detectable PISN rate, and so our predictions here are largely insensitive to the underlying Pop III star formation physics. We finally emphasize that these optimistic predictions also rely crucially on the simplifying assumption that the Pop III IMF remains unchanged between molecular cooling minihalos and pristine atomic cooling halos. Therefore, with all the caveats in mind, these predictions ought to be considered an optimistic upper limit on the Pop III PISN rate that we expect to detect with these next generation survey instruments.

\begin{figure}
    \centering
    \includegraphics[width=0.9\linewidth]{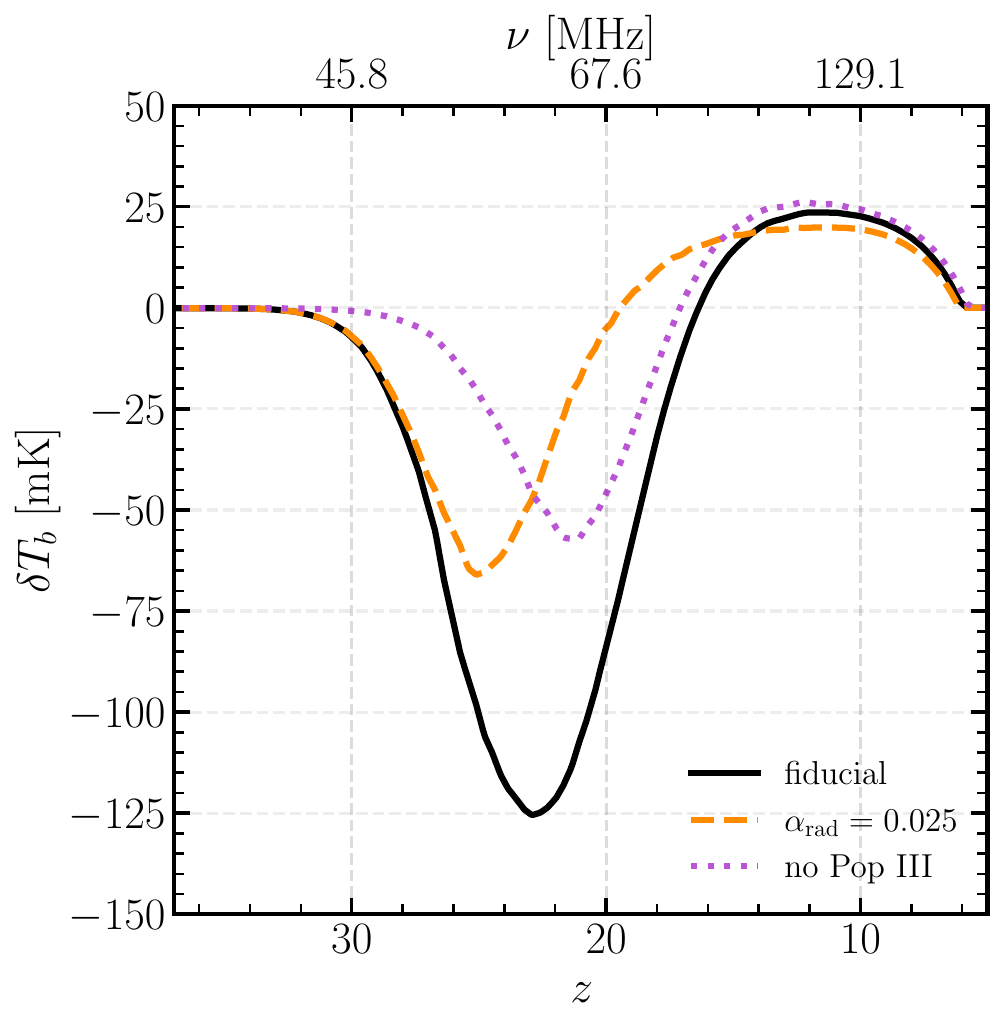}
    \caption{\textbf{The early stages of the 21-cm global signal are highly sensitive to the details of Pop III star formation.} Evolution of the differential brightness temperature of the 21-cm global signal (relative to the CMB) for models with different assumptions for the Pop III physics (different linestyles and colors).}
    \label{fig:dTb_evolution}
\end{figure}

\subsection{The 21-cm global signal}\label{sec:21cm}
Another observation that has been proposed to probe the sites of star formation out to the highest redshifts is the 21-cm signal from neutral hydrogen (see e.g., \citealp{mirocha_unique_2018, mebane_effects_2020, magg_effect_2022, gessey-jones_impact_2022, gessey-jones_signatures_2023, hegde_self-consistent_2023, cruz_effective_2025, gessey-jones_determination_2025, ventura_semi_2025}, for some recent predictions). The strength of the 21-cm emission is determined by a competition between scattering of CMB photons in the IGM, collisions, and scattering of Ly$\alpha$ photons produced by the first stars and galaxies.  

With this model and our tracking of the average thermal and ionization state of the IGM, we can straightforwardly estimate the evolution of the sky-averaged 21-cm brightness temperature, or `global signal,' following the procedure summarized in Section 6.2 of \cite{hegde_self-consistent_2023}. In Figure~\ref{fig:dTb_evolution}, we show this evolution for a few different model variations, which make evident the contribution of early generations of star formation to the thermal state of the IGM. In particular, the inclusion of Pop III stars into a calculation of the IGM's thermal and ionization state leads to a significantly earlier onset and widening of the absorption trough compared to a model with no Pop III stars. Similarly, a model which has enhanced early star formation (due, e.g., to a variation in the details of local radiative feedback in the Pop III SFE), results in earlier Ly$\alpha$ coupling, but also earlier IGM heating, translating the nadir of the absorption trough to earlier times and suppressing its depth. At late times, the global signal is relatively resistant to the details of pristine star formation (i.e., there is minimal difference between the fiducial and maximal pristine star formation curves). In this sense, a detection of the 21-cm global signal at cosmic dawn would be a highly constraining and powerful probe of the details of Pop III star formation physics, even on small scales.

\section{Conclusions}\label{sec:conclusion}
In this paper, we introduce the \abcd\ model, a flexible and comprehensive framework to self-consistently model the growth of the first galaxies from the era of the first stars through the end of reionization. While we make simplifying assumptions along the way, this model presents a clear and efficient method to trace Pop III and II star formation through cosmic time and can be leveraged in a number of contexts. Compared to other models in the literature, we emphasize the modularity and simplicity of our model, which can be run in 30 seconds on a laptop computer. The main findings are as follows:
\begin{itemize}
    \item High-$z$ star formation can be simply modelled with a bathtub-style framework that tracks the inflows and outflows of gas between the ISM, CGM, and IGM reservoirs of a galaxy. Expanding this description to include multiple phases of star formation --- enriched and pristine --- provides a flexible way to track the buildup of stellar mass as the galaxy experiences cycles of bursty star formation.
    \item Pristine star formation at high redshifts can be understood in a relatively modular manner, with different physical mechanisms moderating the star formation rate at different times. At early times, radiative feedback determines the SFRD, while at late times the enrichment of the IGM controls the ability of a halo to continue to form zero-metallicity stars. Ubiquitous across our different model variations is the persistence of Pop III star formation to the end of reionization, with a fiducial late time SFRD of $\rho_{\rm SFR}(z=5)\sim {\rm few\ }\times 10^{-4} M_\odot\ {\rm yr^{-1}\ Mpc^{-3}}$.
    \item Enrichment of the IGM has the most significant effect on the ability of a halo to continue to form Pop III stars. In the limit of inefficient enrichment (maximal pristine star formation), Pop III stars form en masse down to $z\sim 6$ with a Pop III/II ratio of $\sim 10\%$. In a more conservative extreme (wherein the IGM is efficiently enriched), Pop III star formation is heavily suppressed, and only occurs at a low level in the most massive systems, with III/II ratios of $\sim 0.1\%$.
    \item Halo mass provides a useful metric to distinguish between different star formation outcomes. Subject to the efficiency of metal-mixing in the IGM, the most massive halos ($m_h\gtrsim 10^{11}M_\odot$) can continue to form Pop III stars out to the end of reionization, albeit at a subdominant level. In such halos, the ratio of Pop III to II SFR is on the order of $10^{-2}-10^{-1}$ and thus Pop III signatures may be present in extremely deep spectra. At lower masses ($m_h\sim 10^{9}-10^{10} M_\odot$), the joint effects of efficient local enrichment and reionization quenching result in low Pop III/II mass ratios. At the lowest masses, halos achieve higher Pop III/II stellar mass ratios (0.01-1), but star formation in these halos is also quenched by heating of the IGM, so a search for Pop III signatures would likely be limited to relic abundances propagated by SN explosions.
    \item Even under optimistic assumptions for the SN light curves, Pop III SNe are unlikely to be detectable with current and planned JWST observing programs, but may be detectable with wide-field surveys carried out with Euclid or Roman. 
    \item The 21-cm global signal is a sensitive probe of Pop III physics and our model provides a flexible input that can be used to inform and evaluate assumptions in forthcoming inference pipelines.
\end{itemize}
While our model is necessarily simplistic in its structure, and thus omits or simplifies certain processes that are likely to affect the detailed evolution of individual halos at high-redshift (such as mergers, evolution of the IMF, metal mixing, etc.), this approach provides a clear method to understand the sensitivity of Pop III star formation to a variety of internal and environmental processes occurring during the first billion years.

Moving forward, these results suggest that Pop III stars may be detectable even into the Epoch of Reionization and beyond, but theoretical uncertainties limit our ability to confidently estimate the SFRD and luminosities/other signatures we expect to see. To this end, further theoretical work, both numerical and analytical, is necessary to tighten our understanding of small scale star formation physics, such as with the effects of radiative feedback, metal mixing, and the stellar IMF, both within a halo and in its environment. In kind, deep observations with JWST can begin to place constraints on these uncertainties as well. Searching for signatures of pristine stars in more massive halos (through diagnostics such as those suggested in \citealp{venditti_hide_2024} and \citealp{rusta_metal_2025}), for example, can help us place limits on the contribution of Pop III stars to luminous galaxy spectra. In lower mass systems, where we expect reionization quenching to be significant, searching for quasar absorption line signatures of the remnants of early star formation can help probe reionization feedback and star formation in the truly molecular cooling regime. Altogether, our new framework will enable us to guide where and how these observations should be designed, and to physically interpret the results of what we see. 

\section*{Acknowledgments}
We acknowledge that the location where this work took place, the University of California, Los Angeles, lies on indigenous land. The Gabrielino/Tongva peoples are the traditional land caretakers of Tovaangar (the Los Angeles basin and So. Channel Islands). 

We thank Andrea Weibel for providing the JWST PANORAMIC UV luminosity density data and Peng Oh and Roy Zhao for helpful conversations. S.H. is supported by the National Science Foundation Graduate Research Fellowship Program under Grant No. DGE-2034835. Any opinions, findings, and conclusions or recommendations expressed in this material are those of the authors and do not necessarily reflect the views of the National Science Foundation. S.H. acknowledges support from the Future Investigators in NASA Earth and Space Science and Technology (FINESST) Grant No. 80NSSC23K1432 and the UCLA Center for Developing Leadership in Science (CDLS) Fellowship. This work was also supported by NASA through award 80NSSC22K0818 and by the National Science Foundation through award AST-2205900. This work has made extensive use of NASA's Astrophysics Data System (\href{http://ui.adsabs.harvard.edu/}{http://ui.adsabs.harvard.edu/}) and the arXiv e-Print service (\href{http://arxiv.org}{http://arxiv.org}), as well as the following softwares: \textsc{matplotlib} \citep{Matplotlib}, \textsc{numpy} \citep{numpy}, \textsc{astropy} \citep{Astropy}, and \textsc{scipy} \citep{Scipy}.

\begin{appendix}
\section{Appendix A: The gas density profile}
\label{app:gas_density}
One of the key inputs needed to compute the critical mass threshold described in \cite{hegde_self-consistent_2023} is an estimate of the halo's gas density profile, as this sets the rate of H$_2$ buildup and gas cooling. \cite{hegde_self-consistent_2023} used a fit to the results of numerical simulations to quantify this; here we estimate the threshold with an argument similar to that outlined in \cite{oh_fossil_2003}. In essence, the temperature of the CMB sets  a floor on the gas temperature in the IGM, which in turn sets an `entropy floor' in the IGM. Because gas accreted onto the halo cannot end up with less entropy than it carried during accretion, this ultimately results in a `core,' or flattening, in the halo's central gas density profile. This central core reduces the effective gas density (compared to `cusp'-y structures that trace the NFW profile) and thus suppresses the cooling rate, which will modulate our critical star formation threshold.

In the following, assuming that the IGM evolves as a monatomic ideal gas ($\gamma = 5/3$), we define the gas entropy as
\begin{equation}
    \label{eq:entropy_def}
    K \equiv \frac{P}{\rho^{5/3}} = \frac{k_B T}{\mu m_p \rho^{2/3}},
\end{equation}
where $\mu = 1.22$ is the mean molecular weight for neutral primordial gas.

After matter-radiation decoupling ($z_d\sim 150$), the IGM cools adiabatically, and thus (in the absence of any other heating) has a minimum temperature $T_{\rm min, IGM}(z) \approx 2.73\ {\rm K}\ (1+z_d)\big[(1+z)/(1+z_d)\big]^2$. This then sets a \textit{redshift-independent} floor on the entropy given by eq.~\ref{eq:entropy_def}:
\begin{equation}
    \label{eq:entropy_floor}
    K_{\rm floor}(z<z_d) \approx 2.18\times 10^{26}\ {\rm cm^2\ erg\ g^{-5/3}}
\end{equation}
This sets a minimum on the entropy profile for gas accreted from the IGM into a halo, which will manifest as a core in the density profile.

\begin{figure}%
    \centering
    \subfloat[\centering Entropy profiles given by eq.~\ref{eq:entropy_prof} with a minimum set by the entropy floor in eq.~\ref{eq:entropy_floor}.]{{\includegraphics[width=0.4\textwidth]{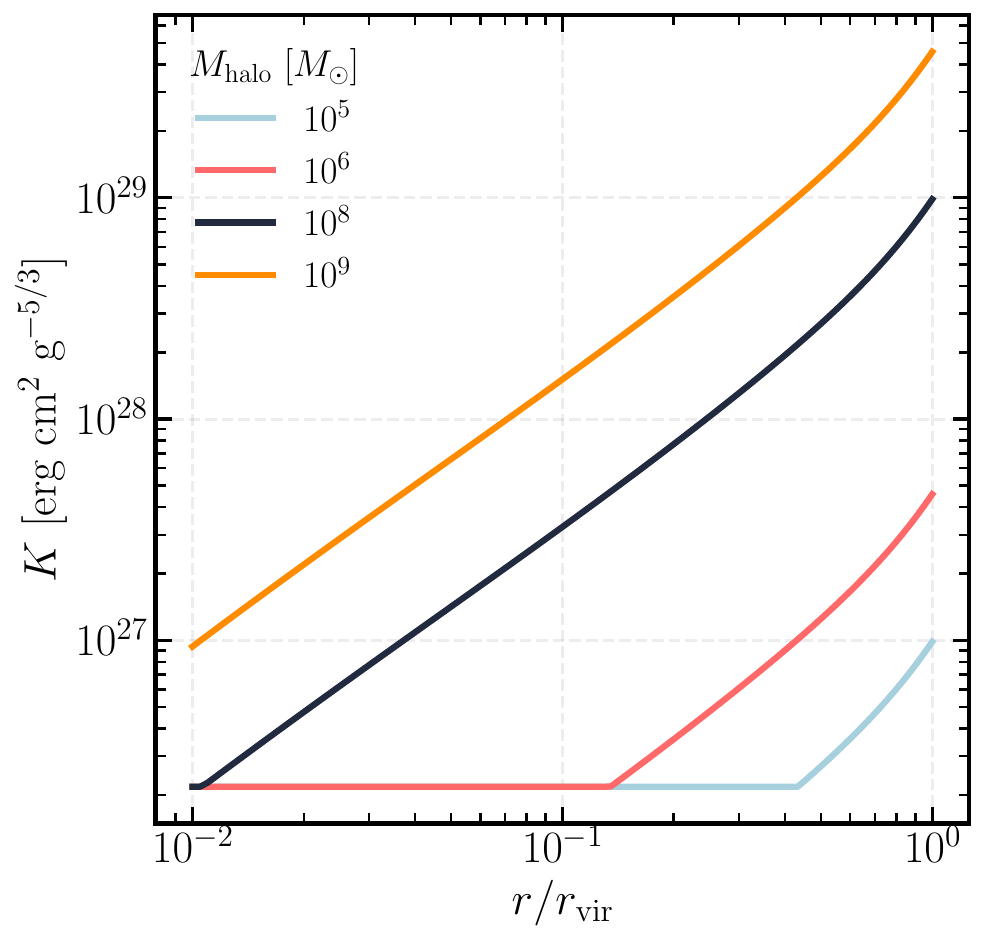} }}%
    \qquad
    \subfloat[\centering The associated density profiles normalized to the mean baryon density.]{{\includegraphics[width=0.4\textwidth]{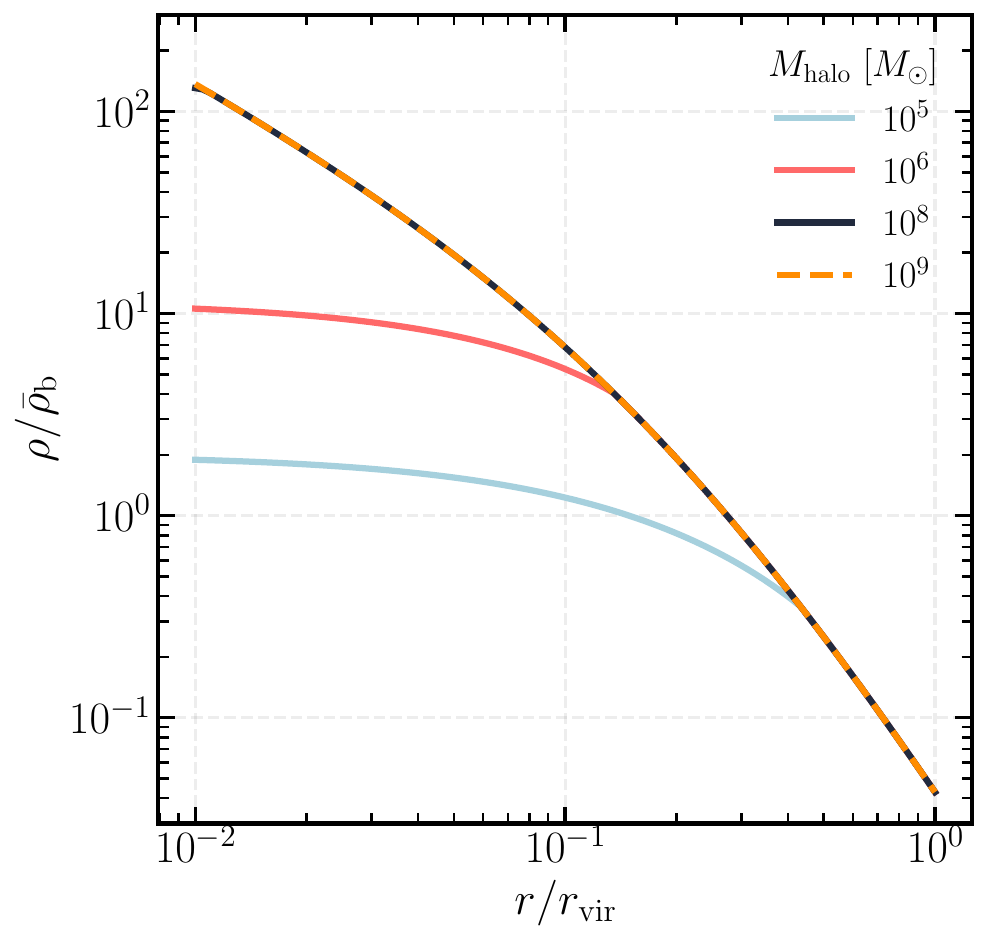} }}%
    \caption{\textbf{The entropy floor introduces a `core' in the central gas density profile.} Entropy and density profiles for four example halo masses (colors).}%
    \label{fig:dens_entropy_profiles}%
\end{figure}

\begin{figure}%
    \centering
    \subfloat[\centering Core densities predicted by eq.~\ref{eq:gas_dens_profile} compared with the results of numerical simulations \citep{oleary_formation_2012}.]{{\includegraphics[width=0.4\textwidth]{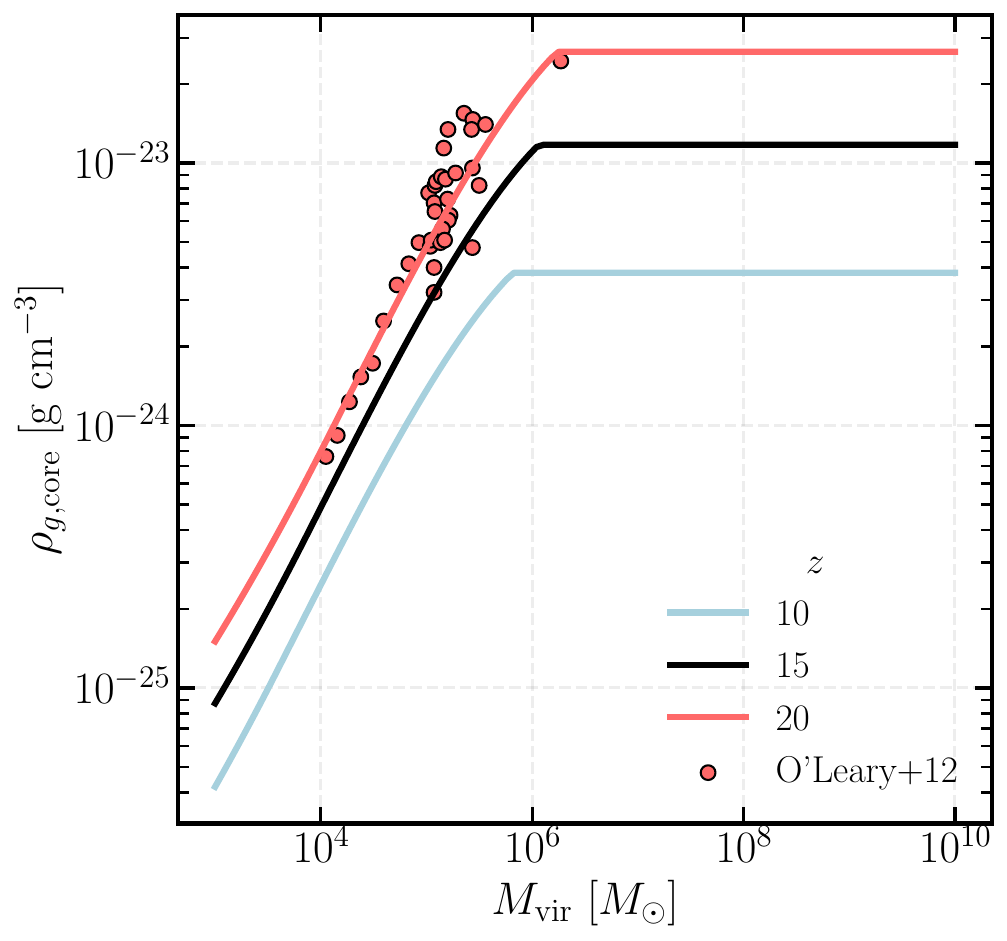} }}%
    \qquad
    \subfloat[\centering Comparison of the central gas number density computed in this work (solid) with the fits used in \cite{hegde_self-consistent_2023} (dashed).]{{\includegraphics[width=0.4\textwidth]{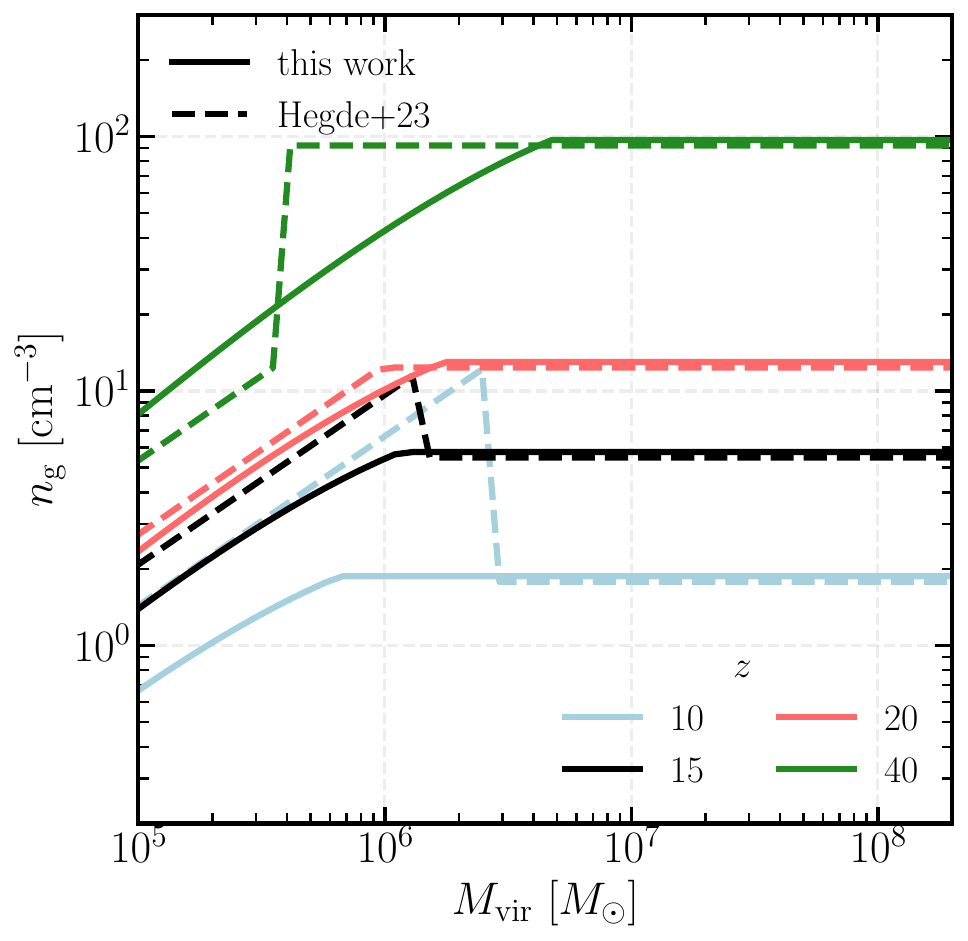} }}%
    \caption{\textbf{The flattening of the central density profile induced by the entropy floor set by the CMB provides excellent agreement with the results of numerical simulations.} Core densities computed at selected redshifts (colors) compared against other estimates from the literature.}%
    \label{fig:dens_profile_comparisons}%
\end{figure}

In the absence of this entropy floor, the density profile of the gas will naively trace that of the dark matter, and thus the entropy profile can be derived from the equation of hydrostatic equilibrium.
\begin{equation}\label{eq:entropy_prof}
    K_{\rm init}(r) = \frac{1}{\rho_g^{5/3}}\Bigg[\frac{\rho_g(r_{\rm vir})k_B T_{\rm vir}}{\mu m_p} - \int_{r_{\rm vir}}^{r}\frac{GM(r')\rho_g(r')}{r'^2}dr'\Bigg],
\end{equation}
where the first term is a boundary term set by the assumption that the gas is heated to the virial temperature at $r_{\rm vir}$ and to compute the initial mass and density profiles, we use the NFW profile \citep{navarro_structure_1996, barkana_beginning_2001}. Combining this with the the entropy floor computed in eq.~\ref{eq:entropy_floor}, the resulting entropy profile is $K(r) = \max\big(K_{\rm floor}, K_{\rm init}(r)\big)$. 

With this modified entropy profile, the hydrostatic equilibrium equation also gives the associated density profile of the gas
\begin{equation}\label{eq:gas_dens_profile}
    \rho_g(r) = \frac{1}{K(r)^{3/5}}\Bigg[\bigg(\frac{\rho_g(r_{\rm vir})k_B T_{\rm vir}}{\mu m_p}\bigg)^{2/5} - \frac{2}{5}\int_{r_{\rm vir}}^r \frac{GM(r')}{K(r')^{3/5}r'^2}dr'\Bigg]^{-5/2},
\end{equation}
where again the first term is set by the boundary conditions at $r_{\rm vir}$. 

Solving eq.~\ref{eq:gas_dens_profile} for a variety of halo masses (which set the normalization of the NFW profile) yields the profiles shown in Figure~\ref{fig:dens_entropy_profiles}. From these profiles, it is clear that the entropy floor indeed flattens the central density profile and the location of that density core moves to smaller radii with increasing halo mass.

Calculating the critical halo mass requires an estimate of the central gas density --- for this, we select $r_{\rm core} \sim 0.1r_{\rm vir}$, consistent with the ISM-CGM boundary chosen throughout the rest of the calculations. In Figure~\ref{fig:dens_profile_comparisons}a, we validate this choice against the results of numerical simulations, finding good agreement with the results of \cite{oleary_formation_2012}. Comparing the results of this calculation with the density estimate used in \cite{hegde_self-consistent_2023} (Figure~\ref{fig:dens_profile_comparisons}b), we see good agreement at high masses, but note differences of a factor of a few at the low-mass end, which manifest as small differences in our calculation of the critical mass.

\section{Appendix B: Luminosity distributions}\label{app:lum_dist}
\begin{figure*}
    \centering
    \includegraphics[width=0.9\linewidth]{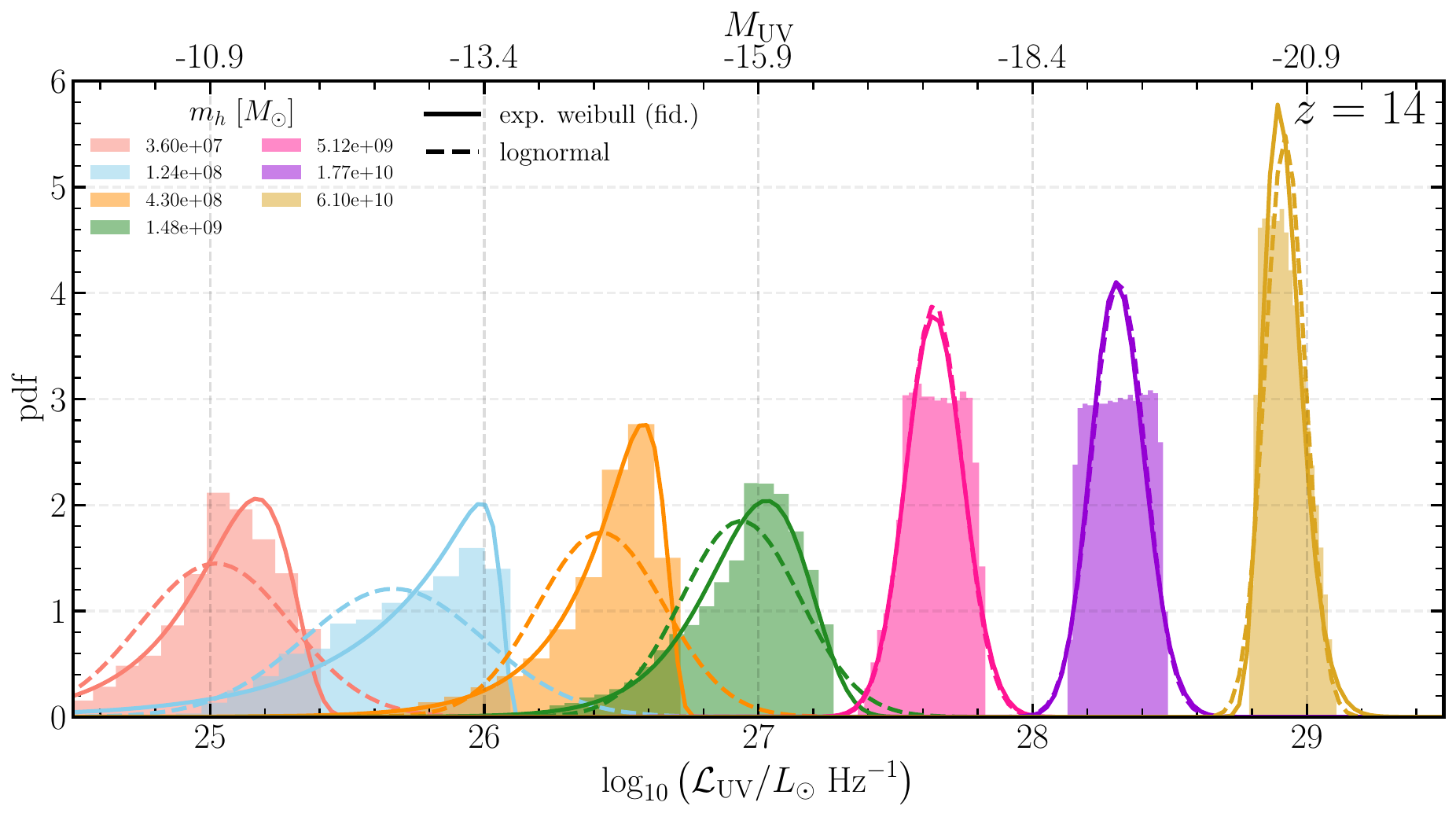}
    \caption{\textbf{The luminosity distributions deviate from lognormal when star formation histories are burstier (at higher redshift and lower halo mass).} UV luminosity distributions in different halo mass bins (colors) at $z=14$ in the fiducial model. The PDFs produced by the best-fit parameters assuming an exponentiated Weibull distribution and a lognormal distribution are shown as solid and dashed lines, respectively.}
    \label{fig:lum_dist}
\end{figure*}

As described in Section~\ref{sec:observables}, the fluctuating star formation histories driven by delayed stellar feedback result in luminosity distributions that deviate from a lognormal approximation. In Figure~\ref{fig:lum_dist}, we show these distributions of UV luminosity as a function of halo mass for the fiducial model at $z\sim 14$. Indeed, we find that for halos smaller than a few $\times 10^{10} M_\odot$, the oscillations in SFR are most pronounced and thus the luminosity distributions are widest, producing an asymmetric distribution skewed left. Because of this, assuming that $P(\log_{10}(\mathcal{L})|m_h)$ for all halo masses (i.e., using the dashed curves to compute the UVLF) leads to an overestimate of the luminosity function in these low mass cases, as is seen in Figure~\ref{fig:UVLF}, for example.

\end{appendix}

\clearpage
\bibliographystyle{mn2e}

\bibliography{main}

\end{document}